\documentstyle[12pt]{article}
\begin{document}

\begin{titlepage}

\hskip10cm {\it  ULB--TH--98/14}

\begin{centering}

\vspace{0.5cm}

\huge{On the quantum Batalin-Vilkovisky formalism and the 
renormalization of non linear symmetries} \\

\vspace{.5cm}

\large{Glenn Barnich$^*$}\\

\vspace{.5cm}

Physique Th\'eorique et Math\'ematique, Universit\'e Libre de 
Bruxelles,
Campus Plaine C.P. 231, B-1050 Bruxelles.

\end{centering}

\vspace{.5cm}

\begin{abstract}
The most convenient tool to study the renormalization of a Lagrangian
field theory invariant under non linear local or global
symmetries is the proper solution to the master equation
of the extended antifield formalism. It is shown that, from the
knowledge of the BRST cohomology, it is possible to
explicitly construct a further extension of the formalism 
containing all the observables of the theory and satisfying 
an extended master equation, with some of the features of
the quantum Batalin-Vilkovisky master equation already present 
at the classical
level. This solution has the remarkable property that all its
infinitesimal deformations can be extended to complete deformations.
The deformed solutions differs from the original one through the
addition of terms related to coupling constant and anticanonical 
field-antifield redefinitions. As a consequence, all theories
admitting an invariant regularization scheme are shown to be
renormalizable while preserving the symmetries, in the sense that both
the subtracted and the effective action satisfy the extended master
equation, and this independently of power counting restrictions. 
The anomalous case is also studied and a suitable definition of the
Batalin-Vilkovisky ``Delta'' operator in the context of dimensional
renormalization is proposed. 
\end{abstract}

\vspace{.5cm}

\footnotesize{$^*$Charg\'e de Recherches du Fonds National
Belge de la Recherche Scientifique.}

\end{titlepage}

\def\be{\begin{eqnarray}}
\def\ee{\end{eqnarray}}
\def\G{\Gamma}
\def\D{\Delta}
\def\qed{\hbox{${\vcenter{\vbox{                         
   \hrule height 0.4pt\hbox{\vrule width 0.4pt height 6pt
   \kern5pt\vrule width 0.4pt}\hrule height 0.4pt}}}$}}
\newtheorem{theorem}{Theorem}
\newtheorem{lemma}{Lemma}
\newtheorem{definition}{Definition}
\newtheorem{corollary}{Corollary}
\newcommand{\proof}[1]{{\bf Proof.} #1~$\qed$}

\section*{Introduction}

The best known example of the renormalization of a theory
invariant under a non linear symmetry is probably non abelian
Yang-Mills theory: on the level of the gauge fixed Faddeev-Popov
action \cite{FaPo}, gauge invariance is expressed through
invariance under the non linear global BRST symmetry
\cite{BRS,Ty}. Some of the crucial points in the analysis are:
(i) the importance of the BRST cohomology as a constraint on the
anomalies and the counterterms of the theory, 
(ii) the anticanonical structure of the theory in terms of
the fields and the sources, to which the BRST variations are coupled,
together with the compact reformulation of all the 
Ward identities in terms of
the Zinn-Justin equation \cite{Zin}, 
and (iii) the insight that BRST exact counterterms can be
absorbed by anticanonical fields and sources redefinitions \cite{Dix}.
The question whether the remaining
counterterms can be absorbed by a redefinition of the coupling
constants of the theory could be settled to the affirmative in the
power counting renormalizable case, through an exhaustive enumeration
of all possible renormalizable interactions \cite{BRS}. 
In the case where one
includes higher dimensional gauge invariant operators, such a property
depends crucially on a conjecture by Kluberg-Stern and Zuber
\cite{KlZu} on the BRST cohomology
in ghost number $0$,
which states that it should be describable by off-shell gauge invariant
operators not involving the ghosts or the sources. This conjecture can be
shown \cite{BBH} to hold in the semi-simple case for which it has been
originally formulated, but its generalization in the presence of abelian
factors is not valid, and this even for power counting renormalizable
theories.
In this last case,
the counterterms violating the generalized Kluberg-Stern and Zuber conjecture 
have been shown to be absent by more involved
arguments from renormalization theory \cite{BBBC}, so that
renormalizability still holds, even if the conjecture does not. 

The classical Batalin-Vilkovisky formalism \cite{BaVi} 
(for reviews, see e.g. 
\cite{HeTe,GPS}) extends the above
techniques to the case of general gauge theories with open gauge algebras
and structure functions, the invariance of the action being expressed
through the central master equation. 
A further extension consists in controlling
at the same time the renormalization of non linear global symmetries
by coupling them with constant ghosts \cite{Bla,BHW}.

A detailed analysis of the
compatibility of the renormalization procedure with invariance
expressed through the master equation has been performed in
\cite{VoTy,VoTy1,VoTy2}, where it has been shown that the renormalized action
is a deformation of the starting point solution to the master
equation. Independently of this result, the fundamental problem of
locality of the construction is raised and 
a locality hypothesis is stated \cite{VoTy}. This hypothesis can be
reinterpretated in a more general framework as the assumption that 
the cohomology of the Koszul-Tate differential \cite{FiHe,Hen}
vanishes in the space of local functionals. While the assumption holds
under certain conditions, which are in particular fulfilled for the
construction of the solution of the master equation, thus guaranteeing
its locality \cite{Hen}, it does not hold in general~; 
the obstructions are related to the non
trivial global currents of the theory \cite{BBH1}, and give rise to
BRST cohomology classes with a non trivial antifield dependence. 

A consequence of this is that there exist observables 
which cannot be made off-shell gauge invariant, even in the case of
closed gauge theories, so that the associated 
deformed solutions of the master equation cannot be
related by a field, antifield and
coupling constant renormalization 
to the starting point solution extended by coupling 
all possible off-shell observables compatible with the symmetries.
 
In \cite{Ans}, renormalization in the context of the
Batalin-Vilkovisky formalism is
reconsidered precisely under the assumption that there are no such
deformations, i.e., 
in the closed case under the analog of the Kluberg-Stern and Zuber
conjecture, and in the open case under the conjecture that all the
BRST cohomology is already contained in the solution to the master equation
coupled with independent coupling
constants\footnote{Note that
it is not true that the 
antifield independent part of the 
cohomology of the differential $(S,\cdot)$ is off-shell
gauge invariant, 
it is in general only weakly
gauge invariant.}, with the conclusion, that the infinities can then
be absorbed by renormalizations.

Finally, in \cite{GoWe} the problem of renormalization under non
linear symmetries is readressed in the context of
effective field theories: it is for instance shown that 
semi-simple Yang-Mills theory and gravity, to which are coupled all possible
(power counting non renormalizable) off-shell observables, 
are such that all the local counterterms needed to
cancel the infinities, can be absorbed through coupling constants,
field and antifield renormalizations, while preserving the symmetry (in
the form of the Batalin-Vilkovisky master equation). 
Theories possessing this last
property, even if an infinite number of coupling constants is needed,
are defined to be renormalizable in the modern sense. 
The difficulty, that is also discussed, is that the non trivial 
infinities are a priori only constrained to belong to 
the BRST cohomology in ghost number $0$, which, because of the non
validity of the generalized Kluberg-Stern and Zuber conjecture 
(taken as an example of a so called structural constraint), does not
guarantee that they can be absorbed by redefinitions of coupling
constants of an action extended by all possible off-shell
observables. What good structural
constraints might be in the general case and if they can be chosen in
such a way as to guarantee renormalizability in the modern sense for
all theories is left as an open question in \cite{GoWe,Wei1}.

A clue to the answer to these questions can be found in
\cite{VoTy,VoTy1,VoTy2}. Indeed, the fact that the divergences 
are such that they
always provide a deformation of the solution of the master equation,
implies in general that the non trivial first order deformations 
satisfy additional cohomological restrictions \cite{BaHe} besides 
belonging to the BRST cohomology. The problem with these
restrictions is that they are non linear in the case of an arbitrary
deformation. Recently \cite{Bar}, it has been shown that the lowest
order additional restriction on the non trivial counterterms is in
fact linear: the counterterms must belong to the kernel of the 
antibracket map, which defines a linear subspace of the BRST cohomology. 

As an (academical) example of how these higher order cohomological 
restrictions work, consider Yang-Mills theories with 
free\footnote{By free, we mean
that the abelian gauge fields have no couplings to 
matter fields,
hence, they have no interactions at all. Their quantization is of
course trivial and we know a priori that no counterterms are
needed.} abelian
gauge fields $A^a_\mu$ as in \cite{GoWe}.
The BRST cohomology in ghost number zero 
contains the term \cite{BBH} 
$$
K=f_{abc}\int d^nx\ F^{a\nu\mu}A^b_\mu A^c_\nu+2A^{*a\mu}A^b_\mu C^c
+C^{*a}C^bC^c,
$$
for completely antisymmetric constants $f_{abc}$, so
that this term is a potential counterterm. At the same time,
the term $k^d\int d^nx\ C^*_d$ belongs to the BRST cohomology
in ghost number $-2$. If we take the action $S_k=S+k^d\int
d^nx\ C^*_d$, we have $1/2(S_k,S_k)=O(k^2)$. This implies
according to the quantum action principle for the regularized theory
that $1/2(\G_k,\G_k)=O(k^2)$ and then, at order $1$ in $\hbar$ for the
divergent part, that $$(S_k,{\G^{(1)}_k}_{div})=O(k^2).$$ The $k$ independent
part of this equation gives the usual condition that the divergent
part of the $k$ independent effective action at first order must be
BRST closed, $(S,{\G^{(1)}}_{div})=0$, and contains in particular the
candidate $K$ above. The $k$ linear part of this equation requires 
$$(\left.\frac{\partial {\G^{(1)}_k}_{div}}{\partial
k^d}\right|_{k=0},S)+(\int d^nx\ C^*_d,{\G^{(1)}}_{div})=0.$$
This condition eliminates the candidate $K$ because $$(\int d^nx\
C^*_d,K)=2f_{abd}\int d^nx\ A^{*a\mu}A^b_\mu+C^{*a}C^b$$ 
is not BRST exact but represents a non trivial BRST
cohomology class in ghost number $-1$. Hence,  
there exists a purely cohomological reason why $K$ cannot
appear as a counterterm. Note that as soon as the abelian fields are
coupled to matter fields, the functionals
$\int d^nx\ C^*_d$ but also $K$ do not belong to the BRST
cohomology any more and the problem with this particular type 
of counterterms
does not arise to begin with.

Another, non trivial example of how antifield dependent counterterms can 
be eliminated by higher order cohomological restrictions is discussed
in the appendix. This example is physically relevant in the
case of the standard model. In the main part of this paper however, 
we will focus on the
general construction of a formalism to deal
with higher order
cohomological restrictions. By general, 
we mean that the construction is independent of the 
concrete gauge theory or type of instability under
consideration and uses only arguments involving the
(integrated, local and antifield dependent) BRST
cohomology and the anticanonical structure of the antifield
formalism. 

A related problem, which is relevant in \cite{VoTy2,Ans,GoWe}, is to
provide a sensible definition of the $\Delta$ operator of 
the quantum Batalin-Vilkovisky master equation \cite{BaVi}. 
Indeed, its expression as a second order functional
differential operator with respect to fields and antifields, obtained
from formal path integral considerations, 
does not make sense when applied to local
functionals. In \cite{TNP,HLW,DPT,PaTr}, the antifield formalism has
been discussed in the context of explicit regularization and 
renormalization schemes and the related question of anomalies
(assumed to be absent in \cite{VoTy2,Ans,GoWe})
has  been adressed. In particular, well defined expressions for the
regularized $\Delta$ operator are proposed at one loop level in
\cite{TNP} in the context of Pauli-Villars regularization
and at higher orders in \cite{PaTr} for non-local regularization. 

The purpose of the present paper is to answer the questions raised in
\cite{GoWe,Wei1} and to show renormalizability in the modern sense for
all gauge theories. We complete the general analysis
of the absorption of divergences in the presence of possibly anomalous
local or global symmetries independently of the form of the BRST
cohomology of the theory. At the same time, a definition of
the quantum Batalin-Vilkovisky $\Delta$
operator in the framework of dimensional renormalization is proposed.

This is done, on the classical level, by
\begin{itemize} 
\item introducing higher order linear maps on the BRST cohomology, 
related to the Lie-Massey brackets \cite{Ret} and constructed by a
perturbative method using constant ghosts and their antifields,
as in the construction \cite{BHW} of the extended antifield formalism,
\item identifying the constant ghosts, coupling all
possible (local integrated) observables of the theory, 
with (generalized) essential coupling constants,
\item showing that the theory to which all observables are coupled 
satisfies an extended master equation at the classical level 
with properties similar to those of the quantum master equation, 
\item proving that the solution to the extended master equation is
complete in the sense that the cohomology associated to
this solution can be obtained by derivation with
respect to the coupling constants,
\item showing how to extend the solution to which an arbitrary cocycle
has been added into a complete deformed solution satisfying the same extended
master equation and the relation of this deformation to 
field-antifield and coupling constant redefinitions.
\end{itemize} 

On the quantum level, in the context of dimensional
regularization in possibly anomalous theories\footnote{As shown in
\cite{HoVe,Ton,Ton1,BrMa,Bon}, dimensional renormalization
can deal consistently with anomalies if 
evanescent terms are taken into account properly.
The same mechanism will be used here
in the context of the Batalin-Vilkovisky formalism to show that the
evanescent breaking terms of the regularized master equation are
responsible for a non trivial $\Delta$ operator, even though 
$''\delta(0)''=0$.}, we show
\begin{itemize} 
\item in the case of an invariant regularization that, order by order,
the divergences are cocycles in ghost number $0$
and as such, they can be absorbed by redefinitions of this solution
determined by coupling constant and anticanonical field-antifield
renormalizations in such a way that both the subtracted and the
effective action satisfy the extended master equation,
\item in the general case of a non invariant scheme, 
that the one loop divergences and anomalies are cocycles in ghost
numbers $0$ and $1$,
\item how the one loop divergences can be absorbed by  
field-antifield and coupling constant redefinitions, 
up to a finite BRST breaking counterterm, chosen in
such a way that only non trivial anomalies appear,
\item how to continue to higher orders by redefinitions preserving the
extended master equation up to BRST breaking counterterms 
in such a way that the anomalous breaking of the extended Zinn-Justin
equation is
entirely determined by the cohomology of the extended BRST
differential in ghost number $1$,
\item the consequences for the definition of the
Batalin-Vilkovisky $\Delta$ operator in dimensional renormalization. 
\end{itemize}

In order to relate the terminology used in this paper with the
one coming from the study of the renormalization of non
abelian Yang-Mills theories \cite{BRS,Zin} (see for instance 
\cite{PiRo,PiSo,Bon1} for reviews), we note that what is called 
extended master equation here corresponds to a generalized
Slavnov-Taylor identity. By generalized, we mean first of all 
the definition of this identity in theories with arbitrary gauge 
structure as proposed by Batalin and Vilkovisky \cite{BaVi}, then
the extension to include the case of (a closed subset of) global
symmetries \cite{Bla,BHW} and finally, the extension proposed here to
include all the generalized observables of the theory.
What is called (local) BRST
cohomology corresponds to the cohomology of
the generalized linearized Slavnov-Taylor operator $S_L$
acting in the space of (integrated) polynomials in the fields, the
sources and their derivatives.  

The main, purely classical part of the paper adresses
the question of how to define stability\footnote{Stability is
defined for instance in \cite{Bon1} as ``the dimension of the
cohomology space of the $S_L$ operator in the Faddeev-Popov neutral
sector should be equal to the number of physical parameters of the
classical action''.} in this generalized context and how 
to construct a 
stable classical action 
{\em independently of power counting restrictions}. This
action then serves as a starting point for the
quantum theory. This is the
translation, in the language of algebraic renormalization, of showing
what is called {\em renormalizability in the modern sense} for generic
gauge theories by the authors of \cite{GoWe}.  

The investigation relies heavily on the
anticanonical structure of variational theories discovered by 
Zinn-Justin \cite{Zin} in the context of non abelian Yang-Mills 
theories (see also \cite {Dix}) and
reintroduced by Batalin and Vilkovisky in the general context, 
in analogy with the generalized Poisson bracket of the Hamiltonian 
formalism \cite{BFV}. 

The quantum part of the paper first shows that if
the regularization scheme is invariant, the
generalized Slavnov-Taylor identities hold at the quantum level. 
In the cases when
dimensional regularization is not an invariant scheme for the
symmetries under considerations, it is shown that BRST breaking 
counterterms
can be choosen in such a way that the breakings of the generalized 
Slavnov-Taylor identity for the renormalized effective action is 
described to each
order by the cohomology of the Slavnov-Taylor operator in ghost number
$1$. The investigation is then used to derive, in the context of
dimensional renormalization at the level of the regularized 
classical action with all its counterterms, a definition of the
Batalin-Vilkovisky $\Delta$ operator \cite{BaVi}, which is an operator
arising naturally in formal path integral manipulations (see also
\cite{Wit}).

\section{Local BRST cohomology, higher order maps and
deformations} 

\subsection{The (extended) antifield formalism}

The central object of the formalism  is a solution $S$ to the 
classical master equation 
\be
\frac{1}{2}(S,S)=0.\label{g1}
\ee
This solution is obtained by constructing a (possibly reducible)
generating set 
of non trivial Noether identities of the
equations of motion $\delta L_0/\delta\phi^i=0$ associated to the
classical action $S_0=\int d^nx L_0$, as well as generating sets of
non trivial reducibility identities for the Noether identities, of non
trivial reducibility identities of the second stage for the previous 
reducibility identities$\dots$. The original set of fields
$\{\phi^i\}$ is extended to the set $\{\phi^A\}$ by 
introducing in addition (i)
ghost fields  
for the non trivial Noether identities, (ii) ghosts for ghosts
associated to the non trivial reducibility identities 
of Noether identities, (iii) ghosts for ghosts for
ghosts for the second stage non trivial reducibility identities$\dots$,
and (iv) antifields $\phi^*_A$ associated to all of the above fields.
In the following, we will denote the
fields and antifields collectively by $z$.

In the classical theory, the relevant space is the space of local
functionals ${\cal F}^*$ in the fields and antifields. 
Under appropriate vanishing
conditions on the fields, antifields and their derivatives at infinity,  
this space is isomorphic to the space of functions in the fields, 
the antifields and a finite number of their derivatives, 
up to total divergences 
(see e.g. \cite{BBH1} and references therein for more details). 
It is an odd graded Lie
algebra with respect to the antibracket $(\cdot,\cdot)$, 
the grading being given by the ghost number, and the 
antibracket being defined by considering the fields and antifields as
canonically conjugate. Note however that the product of two local
functionals is not a local functional, and that there is no 
direct definition
of the second order $\Delta$ operator \cite{BaVi} in this space.

If one requires the solution $S$ of the master equation (\ref{g1})
to be in ghost number $0$, Grassmann even and proper,
i.e., to contain in addition to the starting point action $S_0$ the
gauge transformations related to the generating set of non trivial
Noether identities as well as the various reducibility identities in a
canonical way, one can show \cite{BaVi,VoTy,FiHe,Hen,HeTe} existence
and locality of this solution, with
uniqueness holding up to canonical field-antifield redefinitions.
The BRST differential is then $s=(S,\cdot)$, so that $({\cal
F}^*,(\cdot,\cdot),s)$ is a graded differential Lie algebra with an inner 
differential.

This construction can be extended to include a closed sub-algebra of non
trivial global symmetries, by coupling their generators with constant
ghosts and introducing constant antifields, if due care is taken of
higher order conservation laws \cite{BHW}. The formalism then allows
to control through the master equation both local
and a subset of global symmetries, or, in the absence of gauge
symmetries, the subset of global symmetries alone.
This extension will be a part of a further extension done in the next
section. Indeed, the generators of the global symmetries 
correspond to local BRST
cohomology classes in negative ghost numbers \cite{BBH1}. Here, we will couple
the local BRST cohomology classes in {\em all} the ghost numbers.

A gauge fixed action with well defined propagators as a starting point
for perturbation theory is obtained by introducing a non minimal
sector, which has trivial cohomology \cite{Hen1}, followed by an
anticanonical field-antifield transformation. The new antifields are
not set to zero, but 
kept as sources in the generating functionals of Green's functions in
order to control the renormalization of the symmetries. In the
considerations below, we will only be interested in aspects related to
the local BRST cohomology. The local BRST cohomology of two 
formulations of the
theory related by anticanonical field-antifield transformations 
are isomorphic. This means in particular that all the
results to be derived are independent from the choice of the gauge
fixation. This is the reason why we will not explicitly 
make the steps corresponding to the gauge fixation, i.e., 
introduction of the non minimal sector and transformation to the gauge
fixed basis, although they are understood in the manipulations 
involving the Green's functions.

\subsection{Higher order maps in BRST cohomology 
from homological perturbation theory}

In this section, we construct a generating functional
$\tilde {\cal S}$ for higher order maps in BRST cohomology 
using homological
perturbation theory \cite{HPT} (see \cite{HeTe} for details on homological
perturbation theory in the context of the Batalin-Vilkovisky
formalism) by adapting the construction of the extended antifield
formalism \cite{BHW}. 

For any functional $A\in {\cal F}^*$, the equation $(S,A)=0$ implies
$A=S_A \lambda^A+(S,B)$, where $\lambda^A$ 
is independent of the fields and anti-fields, but can 
depend on the coupling constants of the theory,
with $S_A \lambda^A+(S,B)=0$ iff 
$\lambda^A=0$. In other
words, we suppose that $[S_A]$ is a basis of $H^*(s,{\cal
F})$ over the ring of functions in the coupling constants.

For each $S_A$ of the above basis,
we introduce a constant ``ghost'' $\xi^A$ and a constant ``antifield''
$ \xi^*_A$ such that $gh\  \xi^A=-gh\ S_A$, $gh\
\xi^*_A=-gh\  \xi^A-1$. We consider the space ${\cal E}$ of functionals 
${\cal A}$ of the form 
\be
{\cal A}=A[\phi,\phi^*,\xi]+ \xi^*_A
\lambda^A( \xi),
\ee
i.e., ${\cal A}$ contains a local functional $A$ which admits in 
addition to the dependence on the coupling constants, a dependence
on the constant ghosts $ \xi^A$, and a non integrated piece linear in
the constant antifields $ \xi^*_A$ depending only on the constant
ghosts (and the coupling constants). 

The differential $\tilde \delta$ is defined by $\tilde \delta A=(S,A)$, 
$ \tilde \delta  \xi^*_A=S_A$, and $\tilde \delta \xi^A=0$.
\begin{lemma}
The cohomology of $\tilde\delta$ is trivial, $H(\tilde\delta,{\cal
E})=0$.
\end{lemma}
\proof{ Indeed, 
$\tilde \delta{\cal A}=0$ gives $(S,A)+S_A\lambda^A=0$, and hence
$\lambda^A=0$, so that ${\cal A}=A
=S_A\mu^A+(S,B)=\tilde \delta(B+ \xi^*_A\mu^A)$.}

We define the resolution degree to be the degree in the ghosts
$\xi^A$, which implies that $\tilde\delta$ is of degree $0$.

The extended antibracket is defined by 
\be
(\cdot,\cdot\tilde)=(\cdot,\cdot)+(\cdot,\cdot)_\xi\nonumber\\
=(\cdot,\cdot)
+\frac{\partial^R}{\partial  \xi^A}\frac{\partial^L}{\partial
 \xi^*_A}
-\frac{\partial^R}{\partial  \xi^*_A}\frac{\partial^L}{\partial
 \xi^A}
\ee
and satisfies the same graded antisymmetry and graded Jacobi identity
as the usual antibracket.
The extended
antibracket has two pieces, the old piece $(\cdot,\cdot)$, 
which is of degree $0$, and
the new piece $(\cdot,\cdot)_\xi$, which is of degree $-1$. 
\begin{theorem}\label{t1}
There exists a solution $\tilde {\cal S}\in {\cal E}$ of ghost number $0$ 
to the master equation 
\be
\frac{1}{2}(\tilde {\cal S},\tilde {\cal S}\tilde)=0.\label{m1}
\ee
with initial condition $\tilde {\cal S}=S+S_A \xi^A+ \dots$, where the
dots denote terms of resolution degree higher or equal to $2$. The
cohomology of the differential $\tilde s=(\tilde {\cal S},\cdot
\tilde)$ in ${\cal E}$ is trivial. 
\end{theorem}
\proof{The proofs are by now standard and follow the lines of
\cite{HeTe}. 
Let
$\tilde {\cal S}^1=S+S_A \xi^A$. Note that $\tilde\delta$ is 
the piece of degree $0$ in $(\tilde {\cal S}^1,\cdot\tilde)$ and that
$({\cal S}_k,\cdot\tilde)$ has no piece in degree $0$ if $k\geq 2$.
Suppose that we have constructed $\tilde {\cal S}^k=S+{\cal
S}_1+\dots+{\cal S}_k$ up to degree $k\geq
1$, with 
\be
\frac{1}{2}(\tilde {\cal S}^k,\tilde {\cal S}^k\tilde)={\cal
R}_{k+1}+O(k+2).
\ee
The identity $0=(\tilde {\cal S}^k,\frac{1}{2}(\tilde {\cal S}^k,
\tilde {\cal S}^k\tilde)\tilde)$ then implies, at order $k+1$, that 
$\tilde \delta {\cal R}_{k+1}=0$, so that there exists ${\cal
S}_{k+1}$ such that ${\cal R}_{k+1}+\tilde\delta{\cal
S}_{k+1}=0$. The action $\tilde {\cal S}^{k+1}=\tilde {\cal S}^k+{\cal
S}_{k+1}$ then satisfies 
\be
\frac{1}{2}(\tilde {\cal S}^{k+1},\tilde {\cal S}^{k+1}\tilde)
={\cal R}_{k+1}+\tilde\delta{\cal S}_{k+1}+O(k+2)=O(k+2),
\ee
so that the construction can be continued recursively.

For the second part of the theorem, we develop a cocycle ${\cal A}$
according to the resolution degree, ${\cal A}={\cal A}_M+{\cal
A}_{M+1}+\dots$, with $M\geq 0$. At lowest order the condition 
$(\tilde {\cal S},{\cal A}\tilde)=0$ implies $\tilde \delta {\cal
A}_M=0$ which gives ${\cal A}_M=\tilde \delta {\cal B}_M$ for some
${\cal B}_M$. The cocycle ${\cal A}-(\tilde {\cal S},{\cal
B}_M\tilde)$ is equivalent to ${\cal A}$, but starts at order
$M+1$. Going on in the same way, one can absorb all the terms so that 
${\cal A}=(\tilde {\cal S},{\cal B}\tilde)$ for some ${\cal B}\in{\cal
E}$.  
}

The solution $\tilde {\cal S}$ is of the form 
\be
\tilde {\cal S}=S+\sum_{k=1} S_{A_1\dots
A_k}\xi^{A_1}\dots
\xi^{A_k}+\sum_{m=2}\xi^*_B f^B_{A_1\dots A_m}
\xi^{A_1}\dots \xi^{A_m},
\ee
which implies the graded symmetry of the generalized structure
constants $f^B_{A_1\dots A_m}$ and the functionals $S_{A_1\dots
A_k}$. 
The $\xi^*_A$ independent part of the master equation (\ref{m1})
gives, at resolution degree $r\geq 1$, the relations
\be
(S,S_{A_1\dots A_r})+\sum_{k=1}^{r-1}\frac{1}{2}(S_{(A_1\dots
A_k},S_{A_{k+1}\dots A_r)})(-)^{(A_1+\dots+A_k)(A_{k+1}\dots A_r +1)}
\nonumber\\
+\sum_{k=1}^{r-1}kS_{(A_1\dots A_{k-1}|B|}f^B_{A_{k}\dots
A_r)}=0,\label{m2} 
\ee
where $(\ )$ denotes graded symmetrization.
The first relations read explicitly
\be
(S,S_{A_1})=0,\\
(S,S_{A_1A_2})+\frac{1}{2}(S_{A_1},S_{A_2})(-)^{A_1(A_2+1)}
+S_Bf^B_{A_1A_2}=0,\\
(S,S_{A_1A_2A_3})+(S_{(A_1},S_{A_2A_3)})(-)^{A_1(A_2+A_3+1)}\nonumber\\
+S_Bf^B_{A_1A_2A_3}+2S_{(A_1|B|}f^B_{A_2A_3)}=0,\\
\vdots.\nonumber
\ee
The $\xi^*_A$ dependent part of the master equation (\ref{m1}) gives,
for $r\geq 3$, 
the generalized Jacobi identities 
\be
\sum_{m=2}^{r-1}mf^C_{(A_1\dots A_{m-1}|B|}f^B_{A_m\dots A_r)}=0,
\label{j2}
\ee
the first identities being
\be
2f^C_{(A_1|B|}f^B_{A_2A_3)}=0,\\
2f^C_{(A_1|B|}f^B_{A_2A_3A_4)}+3f^C_{(A_1A_2|B|}f^B_{A_3A_4)}=0,\\
\vdots\nonumber.
\ee

The above solution $\tilde {\cal S}$ is not unique. For a given
initial condition, there is at each
stage of the construction of $\tilde S$, for $k\geq 2$, the 
liberty to add the exact term $\tilde \delta K_k$ to $\tilde {\cal
S}_k$. While this liberty will not affect the structure constants of
order $k$, since a $\tilde\delta$ exact term does not involve a
$\xi^*$ dependent term, it will in general affect the structure
constants of order strictly higher than $k$. 

Furthermore, 
there is a freedom in the choice of the initial condition:
instead of $S_1=S_A\xi^A$, one could have chosen 
$S^\prime_1=\sigma_A^BS_B\xi^A+(S,K_A)\xi^A$ with an invertible matrix
$\sigma_A^B$. If we consider the following anticanonical 
redefinitions:
\be
z^\prime=\exp (\cdot,K_A\xi^A)z,\\
{\xi^\prime}^B=\sigma_A^B\xi^A,\
{\xi^\prime}^*_B={\sigma^{-1}}_B^A\xi^*_A, 
\ee
we have that
$S+S^\prime_1=S(z^\prime)+S_B(z^\prime){\xi^\prime}^B+O(\xi^2)$. 
We can then consider the solution $\tilde {\cal S}^\prime$ in terms of
the new variables. This is equivalent to taking as initial condition 
$S(z^\prime)+S_B(z^\prime){\xi^\prime}^B$ and making the same choices
for the terms of degree higher than $2$ in the new variables
than we did before in the old variables. It is thus always 
possible to make
the choices in the construction of $\tilde {\cal S}$ for $k\geq 2$ in
such a way that the structure constants
$f^B_{A_1\dots A_m}$ do not depend on the choice
of representatives for the cohomology classes and 
transform tensorially with respect to a change of basis in
$H^*(s,{\cal F})$. 

Hence, we have shown 
\begin{theorem}
Associated to a solution $\tilde {\cal S}$ of the
master equation (\ref{m1}),  
there exist multi-linear, graded symmetric
maps in cohomology, defined through the structure constants
$f^B_{A_{1}\dots A_r}$:
\be
l_r:\wedge^r H^*(s)\longrightarrow H^*(s)\\
l_r([S_{A_1}],\dots,[S_{A_r}])=-r[S_{B}]f^B_{A_{1}\dots A_r}\label{gl2}
\ee
\end{theorem}
{\bf Remark:}
For a given $r\geq 2$, let us suppose that the
$[S_{A_1}],\dots,[S_{A_r}]$ are such that the structure constants with
strictly less than $r$ indices vanish, for all choices of $A_i$'s. 
From the identity (\ref{m2}),
it then follows that 
\be
\sum_{k=1}^{r-1}\frac{1}{2}(S_{(A_1\dots
A_k},S_{A_{k+1}\dots A_r)})(-)^{(A_1+\dots+A_k)(A_{k+1}\dots A_r +1)}
\nonumber\\=-rS_{B}f^B_{A_{1}\dots A_r}-(S,S_{A_1\dots A_r}).
\ee
We thus see that, under the above assumption, 
\be
l_r([S_{A_1}],\dots,[S_{A_r}])=[\sum_{k=1}^{r-1}\frac{1}{2}(S_{(A_1\dots
A_k},S_{A_{k+1}\dots A_r)})(-)^{(A_1+\dots+A_k)(A_{k+1}\dots A_r +1)}].
\ee
By comparing with the invariant definitions in \cite{Ret}, 
we identify the maps
$l_r$, under the above assumption,  
as the value of the $r$-place Lie-Massey product 
$[[S_{A_1}],\dots,[S_{A_r}]]$ for the defining system 
$\{S_{A_{i_1}\dots A_{i_k}},\ k=1,\dots,r-1,\   
1\leq i_1<\dots<i_k\leq r\}$.
\qed

\subsection{Coupling constants}

Differentiating the master equation (\ref{g1}) with respect to a coupling
constant $g$, implies that $(S,\frac{\partial^R S}{\partial
g})=0$, so that $[\frac{\partial^R S}{\partial
g}]\in H^0(s)$.

Let us adapt the considerations in \cite{Wei2} to the present context.
\begin{definition} 
A set of coupling constants $g^i$ is essential iff 
the relation $\frac{\partial^R S}{\partial g^i}\lambda^i=(S,\Xi)$
implies $\lambda^i=0$, where $\lambda^i$ may depend on all the
couplings of the theory.
\end{definition}
It follows that for essential couplings the 
$[\frac{\partial^R S}{\partial g^i}]$ are linearly independent in
$H^0(s)$ and  that essential couplings stay
essential after anticanonical field-antifield redefinitions. 

In the following, we suppose that $S$ depends only on essential
couplings. Note that this can always be achieved.
If among the couplings, there is $g$ such that 
$\frac{\partial^R S}{\partial g}
=\frac{\partial^R S}{\partial g^{\bar i}}\lambda^{\bar i}+(S,\Xi)$,
one can show that the dependence of $S$ on
$g$ can be absorbed by an anticanonical, $g$ dependent field-antifield
redefinition and a $g$ dependent redefinition of the other coupling
constants $g^{\bar i}$. 

Since the $[\frac{\partial^R S}{\partial g^i}]$ are linearly
independent, one can construct a basis $[S_A]$ of $H^*(s)$ such that
the $[\frac{\partial^R S}{\partial g^i}]$ are the first elements. Let
us denote the remaining elements by $[S_\alpha]$,
$\{[S_A]\}=\{[\frac{\partial^R S}{\partial g^i}],[S_\alpha]\}$. 
The construction of the generating functional $\tilde {\cal S}$ then
starts with $S(g^i)+\frac{\partial^R S}{\partial
g^i}\xi^i+S_\alpha\xi^\alpha$. 

Consider the action $\bar S=S(g^i+\xi^i)$. A basis of the cohomology
of $\bar S$ is given by $\{[\frac{\partial^R \bar S}{\partial
g^i}],[\bar S_\alpha]\}$, with associated differential
$\bar{\tilde\delta}=(\bar S,\cdot)+\frac{\partial^R \bar S}{\partial
\xi^i}\frac{\partial^L}{\partial \xi^*_i}+\bar
S_\alpha\frac{\partial^L}{\partial  \xi^*_\alpha}$, which is acyclic
in the space where the only dependence on $\xi^i$ is through the
combination $g^i+\xi^i$.
If we take as starting point the action $\bar S+\bar S_\alpha\xi^\alpha$ and
start the perturbative construction of the solution of the master
equation, with resolution degree the degree in the ghost $\xi^\alpha$
alone, the ghosts $\xi^i$ only appear through the combination $g^i+\xi^i$, 
because of the properties of $\bar{\tilde\delta}$.
The solution $\bar{\tilde {\cal S}}$ will then be of
the form 
\be
\bar{\tilde {\cal S}}=\bar S+\sum_{k=1}\bar
S_{\alpha_1\dots\alpha_k}\xi^{\alpha_1}\dots\xi^{\alpha_k}
+\sum_{m=2}(\xi^*_\beta \bar f^\beta_{\alpha_1\dots\alpha_m}+\xi^*_i
\bar f^i_{\alpha_1\dots\alpha_m})\xi^{\alpha_1}\dots\xi^{\alpha_m},
\label{sol1}
\ee
where the $\bar S_{\alpha_1\dots\alpha_k}$, 
$\bar f^i_{\alpha_1\dots\alpha_m}$, $\bar f^i_{\alpha_1\dots\alpha_m}$
depend on the combination $g^i+\xi^i$. 

Now, the solution $\bar{\tilde {\cal S}}$ satisfies the initial
condition $\bar{\tilde {\cal S}}^1=S(g)+\frac{\partial^R S}{\partial
g^i}\xi^i+S_\alpha\xi^\alpha$ in the old resolution degree 
and the master equation
(\ref{m1}). 
We can then  derive the
higher order maps $l_r$ from the solution (\ref{sol1}) and get 
\be
l_r([\frac{\partial^R S}
{\partial g^{i_1}}],\dots,[\frac{\partial^R S}
{\partial g^{i_n}}],[S_{\alpha_{n+1}}],\dots,[S_{\alpha_r}])
\nonumber\\=-r[S_\beta]
\frac{{\partial^R}^n \bar f^\beta_{\alpha_{n+1}\dots\alpha_r}}{\partial
g^{i_n}\dots\partial g^{i_1}} (g)-r[\frac{\partial^R S}
{\partial g^{j}}]
\frac{{\partial^R}^n \bar f^j_{\alpha_{n+1}\dots\alpha_r}}{\partial
g^{i_n}\dots\partial g^{i_1}} (g).\label{exp1}
\ee

In the following, we will make the redefinition
$g^i+\xi^i\longrightarrow \xi^i$, and identify the essential
couplings with some of the constant ghosts. One could of course have
done the converse, i.e., identify some of the constant ghosts with the
essential couplings, but since the anticanonical structure between
the constant ghosts and their constant antifields 
turns out to be crucial, we prefer to do
the former. The remaining constant ghosts can then be considered as 
generalized essential coupling constants since they couple the
remaining BRST cohomology classes, which play the role of generalized
observables in this formalism.

\subsection{Decomposition of $\tilde s$}

The space ${\cal E}$ admits the direct sum decomposition ${\cal
E}=F\oplus G$, where $F={\cal E}|_{\xi^*=0}$ is the space 
of functionals in the field and
antifields with $\xi$ dependence, but no $\xi^*$ dependence, while $G$
is the space of power series in $\xi$ with a linear $\xi^*$ dependence.

The differential $\tilde s$ in ${\cal E}$ induces two
well-defined differentials, $\bar s$ in $F$ and $s_Q$ in $G$ given
explicitly by 
\be 
\bar s=(S(\xi),\cdot)-(-)^{D+1}f^D\frac{\partial^L
}{\partial\xi^D}
\ee 
and 
\be
s_Q=(Q,\cdot)_\xi, 
\ Q=\xi^*_Cf^C(\xi).
\ee
Indeed, for ${\cal
A}=A(\xi)+\xi^*_D\lambda^D(\xi)$, the master equation (\ref{m1})
implies $(\tilde{\cal S},(\tilde{\cal S},{\cal A}\tilde)\tilde)=0$ and
hence $(\tilde{\cal S},\bar s A(\xi)+s_Q
\xi^*_D\lambda^D(\xi)\tilde)=0$ and then $(\bar s)^2A(\xi)+(s_Q)^2
\xi^*_D\lambda^D(\xi)=0$, which splits into two equations because the
decomposition of ${\cal E}$ is direct. 
\begin{theorem}\label{t3}
The cohomology groups $H(\bar s,F)$ and $H(s_Q,G)$ are isomorphic. 
\end{theorem}
\proof{
Let us take $\tilde S=S(\xi)+\xi^*_Cf^C(\xi)$, ${\cal
A}=A(\xi)+\xi^*_C\lambda^C(\xi)$ and ${\cal
B}=B(\xi)+\xi^*_C\mu^C(\xi)$. The extended master equation (\ref{m1})
can be written compactly as 
\begin{eqnarray}
\frac{1}{2}(S(\xi),S(\xi))+\frac{\partial^R
S(\xi)}{\partial\xi^C}f^C=0,
\label{j0}\\
\frac{1}{2}(\xi^*_Cf^C(\xi),\xi^*_Df^D(\xi))_\xi=0,\label{j1}
\end{eqnarray}
so that (\ref{j0}) summarizes (\ref{m2}) and (\ref{j1}), which is 
equal to $\frac{1}{2}(Q,Q)_\xi=0$, or explicitly 
$\frac{\partial^R f^D(\xi)}{\partial \xi^C}f^C(\xi)=0$, 
summarizes the generalized Jacobi identities (\ref{j2}).

The triviality of the cohomology of $(\tilde S,\cdot\tilde)$ in the
space ${\cal E}$, i.e., the fact that the general solution to 
$(\tilde S,{\cal A}\tilde)=0$ is ${\cal A}=(\tilde S,{\cal B}\tilde)$,
is expressed explicitly through the fact that the 
general solution to the set of equations 
\begin{eqnarray}
\left\{ \begin{array}{l}(S(\xi),A(\xi))+\frac{\partial^R
S(\xi)}{\partial\xi^C}\lambda^C(\xi) -(-)^{D+1}f^D\frac{\partial^L
A(\xi)}{\partial\xi^D}=0,\\
(\xi^*_Cf^C(\xi),\xi^*_D\lambda^D(\xi))_\xi=0,
\end{array}\right.\label{tr1}
\end{eqnarray}
is given by 
\begin{eqnarray}
\left\{\begin{array}{l}A(\xi)=(S(\xi),B(\xi))+\frac{\partial^R
S(\xi)}{\partial\xi^C}\mu^C(\xi) -(-)^{D+1}f^D\frac{\partial^L
B(\xi)}{\partial\xi^D},\\
\xi^*_E\lambda^E(\xi)=(\xi^*_Cf^C(\xi),\xi^*_D\mu^D(\xi))_\xi.
\end{array}\right.\label{tr2}
\end{eqnarray}

The cocycle condition $\bar s A(\xi)=0$ 
is given by 
\be
(S(\xi),A(\xi))-(-)^{D+1}f^D\frac{\partial^L
A(\xi)}{\partial\xi^D}=0,
\ee
while the coboundary condition $A(\xi)=\bar s B(\xi)$ is given by 
\be
A(\xi)=(S(\xi),B(\xi))-(-)^{D+1}f^D\frac{\partial^L
B(\xi)}{\partial\xi^D}.
\ee
But according to (\ref{tr1}) and (\ref{tr2}), where we put
$\lambda^D(\xi)=0$, we have 
\be
\bar s A(\xi)=0 \Longleftrightarrow A(\xi)=\bar s B(\xi)+\frac{\partial^R
S(\xi)}{\partial\xi^C}\mu^C(\xi),\label{26}
\ee
under the condition
\be
(\xi^*_Cf^C(\xi),\xi^*_D\mu^D(\xi))_\xi=0\label{27}
\ee
on $\mu^D(\xi)$.
The constraint (\ref{27}) requires
$\xi^*_D\mu^D(\xi)$ to be a $s_Q$ cocycle. 

In order to compute the cohomology $H(\bar s,F)$, we have to
analyze when the decomposition in (\ref{26}) is direct, i.e., we have to
analyze when $\frac{\partial^R S(\xi)}{\partial\xi^C}\mu^C(\xi)$ is a
$\bar s$ coboundary. We thus need to solve the equation $\frac{\partial^R
S(\xi)}{\partial\xi^C}\mu^C(\xi)=-\bar s\bar B$ under the condition
$s_Q\xi^*_D\mu^D(\xi)=0$. But this corresponds precisely to equation
(\ref{tr1}), with $A$ replaced by $\bar B$ and $\lambda(\xi)$ by
$\mu(\xi)$, whose general solution according to (\ref{tr2}) is given
by $\bar B=\bar s C(\xi)+\frac{\partial^R
S(\xi)}{\partial\xi^C}\nu^C(\xi)$ and $\xi^*_D\mu^D(\xi)=s_Q
\xi^*_E\nu^E(\xi)$. (In other words, we are using again the triviality
of $H(\tilde s,{\cal E})$.)  

The map $m(\xi^*_D\mu^D(\xi))=\frac{\partial^R
S(\xi)}{\partial\xi^C}\mu^C(\xi)$ maps $s_Q$ cocyles to $\bar s$
cocyles. We have just proved above that it maps $s_Q$ coboundaries to
$\bar s$ coboundaries, so that the map induced by $m$ in cohomology,
\be
m: H(s_Q,G)\longrightarrow H(\bar s,F),\\
m([\xi^*_D\mu^D(\xi)])=[\frac{\partial^R S(\xi)}{\partial\xi^C}\mu^C(\xi)]
\ee
is
well-defined and injective. Surjectivity of the induced map follows
from (\ref{26}).}

{\bf Discussion:}

(i) In order to compare the starting point cohomology $H^*(s,{\cal F})$
with the cohomology $H^*(\bar s,F)$, we can put the additional
couplings $\xi^\alpha$ to zero in (\ref{26}). The cocycle condition
then reduces to the standard cocycle condition of the non extended
formalism, $s A_{\xi^\alpha=0}=0$. The same operation in the general
solution gives $A_{\xi^\alpha=0}=sB_{\xi^\alpha=0}
+\frac{\partial^R S}{\partial \xi^i}\mu^i_{\xi^\alpha=0} 
+S_\alpha\mu^\alpha_{\xi^\alpha=0}$. Contrary to the ordinary $s$
cohomology, the coefficients
$\mu^A_{\xi^\alpha=0}$ are not
free however, but they come from $\mu^A$'s which are constrained to
satisfy the cocyle condition (\ref{27}). In particular, at 
order $1$ in the new couplings $\xi^\alpha$, (\ref{27}) implies that 
$\mu^\alpha_{\xi^\alpha=0}$ is in the kernel of the map $l_2$, 
$f^A_{\beta\alpha}\mu^\alpha_{\xi^\alpha=0}=0$, which is precisely the
condition used in the examples of the introduction to eliminate 
elements of the cohomology of the theory without the additional
couplings. 
We thus see that the cohomology has become smaller through
the introduction of the additional couplings because the extended
differential encodes higher order cohomological restrictions.
 
(ii) At first sight, it might seem a little strange to introduce new
couplings in order to get information on the renormalization of the
theory without these couplings: that it is convenient and extremely
useful to do so was already realized in the original papers
\cite{BRS} on the
subject: the additional (space-time dependent) couplings 
in these papers are just the sources of the BRS transformations, and
can of course be set to zero after renormalization, if one is only
interested in the renormalization of the effective action itself.

(iii) The result (\ref{26}) implies also that the $\bar s$ 
cohomology is
contained completely in the solution $S(\xi)$ and can be obtained from
it by applying $\frac{\partial^R\cdot}{\partial \xi^A}\lambda^A(\xi)$,
where the coefficients $\lambda^A(\xi)$ are constrained to satisfy
(\ref{27}). In this framework, this is what replaces the concept of 
stability as discussed in the introduction. 

\subsection{``Quantum'' Batalin-Vilkovisky formalism on the classical
level and deformations}
If we define 
\be
{\Delta_c}=(-)^{D}f^D\frac{\partial^L}{\partial \xi^D},
\ee
on $F$, the following properties of the quantum
Batalin-Vilkovisky formalism hold in $F$: 
the operator ${\Delta_c}$ is nilpotent, 
\be
{\Delta_c}^2=0,\label{nil}
\ee
(as a consequence of (\ref{j1}) or (\ref{j2}).) Furthermore, 
\be
{\Delta_c}(A(\xi),B(\xi))=
({\Delta_c} A(\xi),B(\xi))+(-)^{|A|+1}(A(\xi),{\Delta_c} B(\xi)). 
\label{der}
\ee

To the standard solution of the master equation $S$ in ${\cal F}$
corresponds in $F$ the solution $S(\xi)$ of the extended
master equation 
\be
\frac{1}{2}(S(\xi),S(\xi))+{\Delta_c} S(\xi)=0,\label{qme}
\ee
(which is just rewriting (\ref{j0}) using the 
definition of ${\Delta_c}$). 
Because 
\be
\bar s= (S(\xi),\cdot)+{\Delta_c},
\ee
the $\bar s$ cohomology corresponds to 
the quantum BRST cohomology $\sigma$ discussed in \cite{Hen1,HeTe}. Theorem
\ref{t3} shows how to compute the quantum BRST cohomology out of 
the standard BRST cohomology and the higher order maps encoded in
$s_Q$. 

In this analogy, putting $\xi=0$ corresponds to the classical
limit $\hbar\longrightarrow 0$ of the quantum Batalin-Vilkovisky
formalism.

Note however that (i) the space $F$ is not directly an algebra,
because the product of two local functionals is not well defined,
contrary to the formal discussion of the quantum Batalin-Vilkovisky
formalism, where one assumes the space to be an algebra, (ii) the
above ``quantum'' Batalin-Vilkovisky formalism is purely classical and
depends only on the BRST cohomology and the higher order maps of the
theory\footnote{Further details on this cohomological approach  
to the quantum Batalin-Vilkovisky formalism will be considered
elsewhere.}. 

We consider now one parameter deformations of the extended
master equation (\ref{qme}), i.e., in the space $F[t]$ of power series in
$t$ with coefficients that belong to $F$, we want to construct
$S_t(\xi)=S(\xi)+tS_1(\xi)+t^2S_2(\xi)+\dots$ 
such that 
\be
\frac{1}{2}(S_t(\xi),S_t(\xi))+{\Delta_c} S_t(\xi)=0.
\ee

A deformation $S_t(\xi)=S(\xi)+tS_1(\xi)$ to first order in $t$, i.e.,
such that
$\frac{1}{2}(S_t(\xi),S_t(\xi))+{\Delta_c} S_t(\xi)=O(t^2)$ is called an
infinitesimal deformation. 
The term 
linear in $t$ of an  infinitesimal deformation, $S_1(\xi)$,
is a cocycle of the extended BRST differential $\bar s$. 
If $S_1(\xi)$ is
a $\bar s$ coboundary, we call the infinitesimal deformation trivial,
while the parts of $S_1(\xi)$  corresponding to the $\bar s$
cohomology are non trivial. 

\begin{theorem}\label{t4}
Every infinitesimal deformation of the solution $S$ to 
the extended master equation
can be extended to a complete deformation $S_t$.
This extension is obtained by (i) performing a 
$t$ dependent anticanonical field-antifield redefinition
$z\rightarrow z^\prime$, by (ii) performing a $t$ 
dependent coupling constant redefinition
$\xi\rightarrow \xi^\prime$, which does not affect $\Delta_c$, 
and (iii) by adding to $S(z^\prime,\xi^\prime)$ a suitable extension
determined by both coupling constant and the field-antifield
redefinition and vanishing whenever the latter does.

Furthermore, the deformed solution 
considered as a function of the new variables 
$S_t(z(z^\prime,\xi^\prime),\xi(\xi^\prime))$ 
satisfies the extended master equation in terms of the new variables 
and the cohomology $H(\bar s^\prime,F^\prime)$ of 
the differential $\bar s^\prime=(S_t,\cdot)_{z^\prime}+\Delta^\prime_c$
in the space $F^\prime$ of functionals depending on
$z^\prime,\xi^\prime$ is isomorphic to the cohomology $H(\bar s,F)$.
\end{theorem}
\proof{Equation (\ref{26}) and (\ref{27}) imply that $S_1(\xi)=\bar s
B+ \frac{\partial^R S(\xi)}{\partial \xi^C}\mu^C(\xi)$ with
$(\xi^*_Df^D(\xi),\xi^*_C\mu^C(\xi))_\xi=0$. In other words, 
$S_1(\xi)=(\tilde S,B(\xi)+\xi^*_C\mu^C(\xi)\tilde)$. In the extended
space ${\cal E}$, with $z^\alpha=(\phi^B,\phi^*_B)$, 
consider the anticanonical transformation 
\be
{z^\prime}^\alpha=\exp t(\cdot, B(\xi)+\xi^*_C\mu^C(\xi)\tilde)\ z^\alpha=
\exp t(\cdot, B(\xi))z^\alpha\nonumber\\=z^\alpha+t(z^\alpha,B(\xi))+O(t^2),
\ee
\be
{\xi^\prime}^A=\exp t(\cdot, B(\xi)+\xi^*_C\mu^C(\xi)\tilde)\ \xi^A=
\exp t(\cdot,\xi^*_C\mu^C(\xi) )_\xi\ \xi^A\nonumber\\=
\xi^A+t\mu^A(\xi)+O(t^2),
\ee
\be
{\xi^\prime_A}^*=\exp t(\cdot, B(\xi)+\xi^*_C\mu^C(\xi)\tilde)\ {\xi^*_A}
\nonumber\\={\xi^*_A}-t\frac{\partial^L
}{\partial \xi^A}(B(\xi)+\xi^*_C\mu^C(\xi))\label{anti}
+O(t^2).
\ee
Note that $z^\prime=z^\prime(z,\xi)$, $\xi^\prime=\xi^\prime(\xi)$ and
${\xi^*_A}^\prime={\xi^*_A}^\prime(z,\xi,\xi^*)=g_A(z,\xi)
+\xi^*_Bg^B_A(\xi)$,
for a function $g_A(z,\xi)=-t\frac{\partial^L B}{\partial
\xi^A}+O(t^2)$ determined by (\ref{anti}) through both $B$ and $\mu$
and a function 
$g^B_A(\xi)=\delta^B_A-t(-)^{A(B+1)}\frac{\partial^L \mu^B}{\partial
\xi^A}+O(t^2) $ determined by (\ref{anti}) through $\mu$ alone. 

The master equation (\ref{m1}) holds in any variables, 
and thus also in terms of the primed variables. If we denote functions in
terms of the new variables by a prime, we get $\frac{1}{2}(\tilde
S^\prime,\tilde S^\prime\tilde)_{z^\prime,\xi^\prime}=0$. Because the
transformation is anticanonical, we also have 
\be
\frac{1}{2}(\tilde
S^\prime,\tilde S^\prime\tilde)_{z,\xi}=0. \label{nspl}
\ee
Since
\be
\tilde S^\prime=S^\prime+g_A{f^\prime}^A+\xi^*_Bg^B_A{f^\prime}^A,
\ee
equation (\ref{nspl}) splits into 
\be
\frac{1}{2}(S^\prime+g_A{f^\prime}^A,S^\prime+g_A{f^\prime}^A)_z
+\frac{\partial^R}{\partial\xi^D}(S^\prime+g_A{f^\prime}^A)
g^D_E{f^\prime}^E=0,\label{45}\\
\frac{1}{2}(\xi^*_Bg^B_A{f^\prime}^A,\xi^*_Dg^D_C{f^\prime}^C)_\xi=0.
\label{nil1}
\ee
We have 
\be
\frac{d
(S^\prime+g_A{f^\prime}^A)}{dt}|_{t=0}
=(S(\xi),B(\xi))+{\Delta_c} B+\frac{\partial^R
S(\xi)}{\partial \xi^D}\mu^D=S_1(\xi),
\ee
and, because $\xi^*_E\mu^E(\xi)$ is a $s_Q$ cocycle, the relation 
\be
g^D_C{f^\prime}^C=f^D.\label{us}
\ee
Indeed, if we consider the above canonical transformation with $B=0$,
i.e., $\exp
t(\cdot,\xi^*\mu)_\xi)$ alone,
$\xi^*_Cg^C_Df^D(\xi^\prime)={\xi^*}^\prime_Df^D(\xi^\prime)=
\exp t(\cdot,\xi^*\mu)_\xi\xi^*_Ef^E=\xi^*_Ef^E$, because 
$(\xi^*_Ef^E,\xi^*_G\mu^G)_\xi=0$.
This shows the first part of the theorem, with
$S_t=S^\prime+g_A{f^\prime}^A$. 

In order to prove the second part, we first note that 
\be
{\Delta_c}^\prime=(-)^Df^D(\xi^\prime)\frac{\partial^L}{\partial
{\xi^\prime}^D}=(-)^Df^D(\xi^\prime)\frac{\partial^L\xi^C}{\partial
{\xi^\prime}^D}\frac{\partial^L}{\partial {\xi^C}}=\Delta_c
\ee
because 
\be
g^C_D=\frac{\partial^L\xi^C}{\partial
{\xi^\prime}^D}.\label{tt}
\ee
Indeed, 
we have
$\delta^A_B=({\xi^\prime}^A,{\xi^*_B}^\prime\tilde)_{z^\prime,\xi^\prime}
=({\xi^\prime}^A,{\xi^*_B}^\prime)_{\xi}
=\frac{\partial^L{\xi^\prime}^A}{\partial\xi^C}g^C_B$. 
Together with (\ref{45}), this implies
\be
\frac{1}{2}(S^\prime+g_A{f^\prime}^A,S^\prime
+g_A{f^\prime}^A)_{z^\prime}+\Delta^\prime_c(S^\prime+g_A{f^\prime}^A)=0.
\ee
We then start from the relations 
\be
(\tilde S^\prime,{\cal
A}^\prime\tilde)_{z^\prime,\xi^\prime}=0\Longleftrightarrow 
{\cal
A}^\prime=(\tilde S^\prime,{\cal B}^\prime\tilde)_{z^\prime,\xi^\prime},
\ee
where ${\cal A}^\prime=A^\prime+{\xi^*}^\prime_A{\lambda^\prime}^A$
and ${\cal B}^\prime=B^\prime+{\xi^*}^\prime_A{\rho^\prime}^A$.
These relations hold with the bracket taken in the old variables, 
because the transformation is
anticanonical. Writing the resulting relations explicitly, 
using (\ref{us}), 
we get that the set of relations 
\be\left\{\begin{array}{l}
(S^\prime+g_A{f^\prime}^A,A^\prime+g_B{\lambda^\prime}^B)+
\frac{\partial^R }{\partial
\xi^C}(S^\prime+g_A{f^\prime}^A)g^C_D{\lambda^\prime}^D \\
+(-)^Af^A\frac{\partial^L}{\partial
\xi^A}(A^\prime+g_B{\lambda^\prime}^B)=0,\\
\xi^*_A\frac{\partial^R g^A_B{f^\prime}^B}{\partial \xi^C}g^C_D
{\lambda^\prime}^D+(-)^Dg^D_C{f^\prime}^C\frac{\partial^L}{\partial
\xi^D}(\xi^*_Ag^A_B{\lambda^\prime}^B)=0,
\end{array}\right.\label{ch1}
\ee
is equivalent to the set 
\be\left\{\begin{array}{l}
A^\prime+g_B{\lambda^\prime}^B=
(S^\prime+g_A{f^\prime}^A,B^\prime+g_C{\rho^\prime}^C)+
\frac{\partial^R }{\partial
\xi^C}(S^\prime+g_A{f^\prime}^A)
g^C_D{\rho^\prime}^C \\+(-)^Af^A\frac{\partial^L}{\partial
\xi^A}(B^\prime+g_C{\rho^\prime}^C),\\
\xi^*_Ag^A_B{\lambda^\prime}^B=
\xi^*_A\frac{\partial^R g^A_B{f^\prime}^B}{\partial \xi^C}g^C_D
{\rho^\prime}^D+(-)^Dg^D_C{f^\prime}^C\frac{\partial^L}{\partial
\xi^D}(\xi^*_Ag^A_B{\rho^\prime}^B).
\end{array}\right.\label{ch2}
\ee
Using (\ref{tt}), the last equations in (\ref{ch1}) and (\ref{ch2})
are just the $s_Q$ cocycle and coboundary conditions, expressed in
terms of the ${\xi^*}^\prime,\xi^\prime$ variables. 
Following the same reasoning as in the proof of theorem \ref{t3}, we
get, 
\be
(S^\prime+g_A{f^\prime}^A,A^\prime)+\Delta_cA^\prime=0\nonumber\\
\Longleftrightarrow
A^\prime=(S^\prime+g_A{f^\prime}^A,B^\prime+g_C{\rho^\prime}^C)
\nonumber\\+\Delta_c(B^\prime+g_C{\rho^\prime}^C)+
\frac{\partial^R }{\partial
{\xi^\prime}^C}(S^\prime+g_A{f^\prime}^A){\rho^\prime}^C,\\
({\xi^*}^\prime_A{f^\prime}^A,
{\xi^*}^\prime_B{\rho^\prime}^B)_{\xi^\prime}=0,
\ee
where $\frac{\partial^R }{\partial
{\xi^\prime}^C}(S^\prime+g_A{f^\prime}^A){\rho^\prime}^C$ is 
$(S^\prime+g_A{f^\prime}^A,\cdot)+\Delta_c$ exact iff 

\noindent ${\xi^*}^\prime_B{\rho^\prime}^B=({\xi^*}^\prime_A{f^\prime}^A,
{\xi^*}^\prime_C{\nu^\prime}^C)_{\xi^\prime}$. Since
$(S^\prime+g_A{f^\prime}^A,\cdot)+\Delta_c=
(S^\prime+g_A{f^\prime}^A,\cdot)_{z^\prime}
+\Delta^\prime_c=\bar s^\prime$, 
we get that
$H(\bar s^\prime,F^\prime)$ is determined by 
$\frac{\partial^R  }{\partial
{\xi^\prime}^C}(S^\prime+g_A{f^\prime}^A){\rho^\prime}^C$ 
corresponding to the class 
$\frac{\partial^R S}{\partial
\xi^C}{\rho}^C$ of $H(\bar s,F)$.}

{\bf Remark:} Note that one can prove in the same way that the relations
$\frac{1}{2}(A,A)+\Delta_c A=C$ and  
$(A,D)+\Delta_c D= E$ become, after the change of variables,
$\frac{1}{2}(A^\prime + g_A 
{f^\prime}^A,A^\prime + g_A 
{f^\prime}^A)+\Delta_c (A^\prime + g_A 
{f^\prime}^A)=C^\prime$, respectively
$(A^\prime + g_A {f^\prime}^A,D^\prime) +\Delta_c D^\prime=E^\prime$.  

\section{Renormalization}

We show how the renormalization can be performed
while respecting as much as possibe the symmetry in the form of the
extended master equation: the
corresponding extended Zinn-Justin equation for the renormalized
effective action is shown to be broken only by non trivial
anomalies. 

\subsection{Regularization}

We apply the discussion and notations 
of dimensional regularization given in \cite{Ton1} to
the extended master equation (\ref{qme}) and its solution 
$S(\xi)$\footnote{It is also possible to apply directly the
rigorously proved renormalized quantum action principles 
\cite{QAP} (proved in the context of dimensional renormalization 
in \cite{BrMa}, see also \cite{Bon}). We choose not to do
so here, because we want to keep the divergent counterterms explicitly
in the discussion, to allow on the one hand for a direct comparison 
with 
the discussions in \cite{Ans,GoWe} and, on the other hand, to be able
to define the Batalin-Vilkovisky $\Delta$ operator in this context.}. 
In
the following we will always understand the $\xi$ dependence without
explicitly indicating it. Local functionals are understood to
belong to $F$. In fact, we will only use the following three 
properties of dimensional regularization \cite{Ton1}: 
\begin{itemize}
\item the regularized action $S_\tau=\sum_{n=0}\tau^n S_n$ is a
polynomial or a power series in $\tau$, the classical 
starting point action $S$ corresponding to $S_0$\footnote{We 
assume that all the algebraic relations holding for the classical
action $S$ hold in the regularized theory for $S_0$.},
\item if the renormalization has been carried out to $n-1$ loops, the
divergences of the effective action at $n$ loops are poles in $\tau$
up to the order $n$ with residues that are local functionals, and
\item the regularized
 quantum action principle holds (see the first reference of 
\cite{BrMa}, sections II.3 and II.4).
\end{itemize}

Let $\theta_\tau=\frac{1}{2\tau}(S_\tau,S_\tau)
+\frac{1}{\tau}{\Delta_c}
S_\tau$. Note that $\theta_\tau$ is of order $\tau^0$ because $S_0$ 
satisfies the extended master
equation. $\theta_\tau$ characterizes the breaking of the extended
master equation due to the regularization. In order to control this
breaking during renormalization, it is useful to couple it with a
global source $\rho^*$ in ghost number $-1$ and consider 
$S_{\rho^*}=S_\tau+\theta_\tau\rho^*$. On the classical, regularized
level, we have, using $(\rho^*)^2=0$, and the properties (\ref{nil})
and (\ref{der}) of ${\Delta_c}$,
\be
\frac{1}{2}(S_{\rho^*},S_{\rho^*})+{\Delta_c}
S_{\rho^*}=\tau\frac{\partial^R S_{\rho^*}}{\partial\rho^*},
\ee
Applying the quantum action principle, we get, for the regularized
generating functional for 1PI irreducible vertex functions 
$\Gamma_{\rho^*}$ associated to $S_{\rho^*}$,
\be
\frac{1}{2}(\G_{\rho^*},\G_{\rho^*})+{\Delta_c}
\G_{\rho^*}=\tau\frac{\partial^R \G_{\rho^*}}{\partial\rho^*},
\ee
which splits, using $(\rho^*)^2=0$, into 
\be
\frac{1}{2}(\G,\G)+{\Delta_c}
\G=\tau\frac{\partial^R \G_{\rho^*}}{\partial\rho^*},\label{endl1}\\
(\G,\frac{\partial^R\G_{\rho^*}}{\partial\rho^*})
+{\Delta_c}\frac{\partial^R\G_{\rho^*}}{\partial\rho^*}=0\label{endl2}.
\ee

\subsection{Invariant regularization}

Before proceeding with the general analysis, let us briefly discuss
the case when dimensional regularization is an invariant
regularization scheme for the symmetries under considerations.
(The arguments below can be adapted in a
straightforward way to other invariant regularization schemes).

In this case, $\theta_\tau$ vanishes and we have 
\be
\frac{1}{2}(S_\tau,S_\tau)+\Delta_c S_\tau=0. 
\ee
For the regularized generating functional, we get 
\be
\frac{1}{2}(\G,\G)+\Delta_c \G=0,\label{ein}
\ee
where by assumption,
$\G=S_\tau+\hbar\sum_{n=-1}\tau^n\G^{(1)n}+O(\hbar^2)$. 
To order $\hbar/\tau$, (\ref{ein}) gives 
\be
\bar s \G^{(1)-1}=0\Longleftrightarrow \G^{(1)-1}=\bar
s{\Xi_1}+\frac{\partial^R S_0}{\partial \xi^A}\mu^A_1,
\ee
where $\bar s=(S_0,\cdot)+\Delta_c$ and $s_Q\ \xi^*_A\mu^A_1=0$.

We then make the following 
change of fields, antifields and
coupling constants:
\be
z^1=\exp -\frac{\hbar}{\tau}(\cdot ,{\Xi_1}+\xi^*\mu_1\tilde)z ,\\
{\xi^1}=\exp -\frac{\hbar}{\tau}(\cdot ,\xi^*\mu_1)_\xi
\xi .
\ee
If we denote by a superscript $1$ functions depending on these new
variables, we have, according to the remark after theorem \ref{t4},
that the action $S_{R1}=S^1_\tau+{g_1}_A{f^1}^A$, where ${g_1}_A$ 
is determined through the generators 
${\Xi_1}$ and $\mu^C_1$ of the first redefinition, 
satisfies the extended master
equation (\ref{qme}),
\be
\frac{1}{2}(S_{R_1},S_{R_1})+\Delta_c S_{R_1}=0,
\ee
and allows to absorb the one loop divergences, since
$S_{R1}=S_\tau-\hbar/\tau\G^{(1)-1}+\hbar O(\tau^0)+O(\hbar^2)$.
We thus have for the corresponding regularized generating functional 
$\G_{R_1}=S_\tau+\hbar\sum_{n=0}\tau^n\G_{R_1}^{(1)n}
+\hbar^2\sum_{n=-2}\tau^n
\G_{R_1}^{(2)n}+O(\hbar^2)$,
\be
\frac{1}{2}(\G_{R_1},\G_{R_1})+\Delta_c \G_{R_1}=0.
\ee
At order $\hbar^2/\tau^2$ , we get 
\be
\bar s \G_{R_1}^{(2)-2}=0\Longleftrightarrow \G^{(2)-2}=\bar
s\Xi_{2,-2}+\frac{\partial^R S_0}{\partial \xi^A}\mu^A_{2,-2},
\ee
with $s_Q\ \xi^*_A\mu^A_{2,-2}=0$.
The appropriate change of variables is
\be
z^{2,-2}=\exp -\frac{\hbar^2}{\tau^2}(\cdot
,{\Xi_{2,-2}}+{\xi^*}\mu_{2,-2}\tilde) z ,\\
{\xi^{2,-2}}=\exp -\frac{\hbar^2}{\tau^2}(\cdot ,
{\xi^*}\mu_{2,-2})_\xi\xi.
\ee
The regularized action 
$S_{R_{2,-2}}=S_{R_1}^{2,-2}+{g_{2,-2}}_A{f^{2,-2}}^A$ satisfies the
extended master equation and allows to absorb the poles
of order $\hbar^2/\tau^2$:
\be
\frac{1}{2}(S_{R_{2,-2}},S_{R_{2,-2}})+\Delta_c S_{R_{2,-2}}=0,
\ee
and 
$\G_{R_{2,-2}}=S_\tau+\hbar\sum_{n=0}\G_{R_{2,-2}}^{(1)n}
+\hbar^2\sum_{n=-1}\tau^n
\G_{R_{2,-2}}^{(2)n}+O(\hbar^3)$.

In the same way, one can then proceed to 
absorb the poles of order $\hbar^2/\tau$ to get a regularized action 
$S_{R_{2,-1}}$ and an associated two loop finite effective action 
$\G_{R_{2,-1}}$, with both actions satisfying the extended master
equation.

Going on recursively to higher orders in $\hbar$, we can achieve, 
through a
succession of redefinitions, the absorptions of the infinities to
arbitrary high order in the loop expansion, while preserving the
extended master equation for the redefined action and the
corresponding generating functional.

\noindent [Symbolically, 
\be
\frac{1}{2}(S_{R_\infty},S_{R_\infty})+\Delta_c S_{R_\infty}=0,
\ee
with $\G_{R_\infty}$ finite and satisfying
\be
\frac{1}{2}(\G_{R_\infty},\G_{R_\infty})+\Delta_c \G_{R_\infty}=0.]
\ee

We have thus shown:
\begin{theorem}
In theories admitting an invariant regularization scheme, the
divergences can be absorbed by successive redefinitions in such a way
that both the subtracted and the effective action satisfy the extended
master equation. 
\end{theorem}

\subsection{Structural constraints and cohomology of $\bar s$}

Structural constraints have been introduced in \cite{GoWe} to give in
particular cases a sufficient, but not a necessary condition for
renormalizability in the modern sense. In the cases of semi-simple
Yang-Mills theories or gravity for instance, these constraints 
correspond to the prescription of a choice for the representatives 
of the BRST cohomology classes in ghost number $0$ (to be coupled to the bare
action, if not already contained therein): the representatives should
be taken to be independent of the antifields. Because one can prove that in
every BRST cohomology class in ghost number $0$, there exists such a 
representative, renormalizability in the modern sense is
guaranteed to hold. 

But in these examples, one expects renormalizability in the modern
sense to hold, even if one chooses different representatives for the
cohomology classes. 
Consider for instance semi-simple Yang-Mills theory. The choice of
representatives in agreement with the structural constraint is
to take the Yang-Mills action itself as a representative, or, in other
words, to consider the coupling $k$ in front of the Yang-Mills action as an
essential one. One could
also take the derivative of the solution of the master equation with
respect to the coupling constant $g$ associated with the structure
constant as a representative for this cohomology class, and this
representative depends on the antifields. That these two terms are in
the same cohomology class follows from the well-known field
redefinitions that allow to absorb either one of them. 
In other words, only one of the couplings $k$ or $g$ is an essential
one. Choosing $g$ as an essential coupling does not respect the
structural constraint, but clearly, one does not expect the validity of
renormalizability in the modern sense to depend on this choice.

What has been shown in the previous sections is that renormalizability
in the modern sense does not depend on how one chooses the
representatives, or equivalently, the essential couplings, and that 
structural constraints are not necesseary conditions for
renormalizability in the modern sense. This has been done by taking
into account higher order cohomology restrictions, incorporated in the
extended antifield formalism through 
the cohomology of the operator $\bar s$. 
As shown in \cite{GoWe},
structural constraints are nevertheless very useful in concrete cases,
to show renormalizability in the modern sense,  
without using 
the more heavy machinery developped here, which consists in
controling the renormalization of the complete theory with
all its generalized observables. 

In the non anomalous case, the extended master equation for the
effective action implies a remarkable stability of the quantum theory:
while the expression of the generalized observables of the theory are
affected by quantum corrections, their antibracket algebra stays the
same than in the classical theory. In particular, the usual
algebra of the generators of the global symmetries (whether linear or
not) is the same in the classical and the quantum theory\footnote{The
author is grateful to F. Brandt for pointing this out.}. This is
because the antibracket algebra of the BRST cohomology classes in
negative ghost numbers just reflects the ordinary algebra of the
symmetries they represent.  

\subsection{One loop divergences and anomalies}

Let us now go back to the general case where the dimensional
regularization scheme is not invariant\footnote{The derivation of some
of the results below in the framework of algebraic renormalization
\cite{PiSo} will be discussed elsewhere \cite{Bar2}.}.

At one loop, we get from (\ref{endl1}) and (\ref{endl2}) 
\be
(S_\tau,\G^{(1)})+{\Delta_c} \G^{(1)}=\tau\theta^{(1)},\label{1lc}\\
(S_\tau,\theta^{(1)})+(\G^{(1)},\theta_\tau)+{\Delta_c}\theta^{(1)}=0,
\label{1la} 
\ee
where $\G^{(1)}$ and $\theta^{(1)}$ are respectively the one loop
contributions of $\G$ and
$\frac{\partial^R\G_{\rho^*}}{\partial\rho^*}$. By assumption, we have
both 
$\G^{(1)}=\sum_{n=-1}\tau^n\G^{(1)n}$ and $\theta^{(1)}
=\sum_{n=-1}\tau^n\theta^{(1)n}$, where $\G^{(1)-1},\theta^{(1)-1}$
are local functionals. 

At $\frac{1}{\tau}$, equation (\ref{1lc}) gives
\be
\bar s \G^{(1)-1}=0,
\ee
Using this equation together with $\theta_0=\bar s S_1$, equation
(\ref{1la}) implies
\be
\bar s (\theta^{(1)-1}-(\G^{(1)-1},S_1))=0.
\ee
Equation (\ref{1lc}) also gives at order $\tau^0$
\be
\bar s \G^{(1)0}=\theta^{(1)-1}-(\G^{(1)-1},S_1),\label{ma}
\ee
which allows us to identify the combination
$A_1=\theta^{(1)-1}-(\G^{(1)-1},S_1)$ as the one loop anomaly and
explicitly shows its locality.
We have thus shown in the case of a non invariant
regularization scheme:
\begin{theorem}
The one loop divergences $\G^{(1)-1}$ and the one loop anomalies $A_1$
are $\bar s$ cocycles in ghost number $0$ and $1$ respectively.
\end{theorem}

\subsection{One loop renormalization}

According to (\ref{26}), we have 
\be
\G^{(1)-1}=\bar s
\Xi_1+\frac{\partial^R S_0}{\partial \xi^D}\mu^D_1
\ee
and 
\be
A_1=\bar s\Sigma_1
+\frac{\partial^R S_0}{\partial \xi^E}\sigma^E_1, \label{ma1}
\ee
with $s_Q\ \xi^*_A \mu^A_1=0=s_Q \xi^*_B\sigma^B_1$.
The appropriate change of variables is now 
\be
z^1=\exp -\frac{\hbar}{\tau}(\cdot ,{\Xi_1}+\xi^*\mu_1\tilde)z ,\\
{\xi^1}=\exp -\frac{\hbar}{\tau}(\cdot ,\xi^*\mu_1\tilde)\xi .
\ee
The renormalized one loop action is 
\be
S_{{R_1}}=S^{1}_\tau+{g_1}_A{f^1}^A-\hbar\tilde\Sigma_1^1=
S_\tau-\frac{\hbar}{\tau}\G^{(1)-1}+\hbar O(\tau^0)
+O(\hbar^2),
\ee
where $\tilde\Sigma_1$ remains to be determined. 
Using the remark after theorem \ref{t4}, we get 
\be
\theta_{R_1}\equiv\frac{1}{2\tau}({S_{{R_1}}},{S_{{R_1}}})
+\frac{1}{\tau}
\Delta_c{S_{{R_1}}}
=\theta_\tau^1-\frac{\hbar}{\tau}(\bar s \tilde\Sigma_1)^1
+O(\hbar^2)\nonumber\\
=\theta_\tau-\frac{\hbar}{\tau}\bar s[\tilde\Sigma_1+(S_1,\Xi_1)
+\frac{\partial^R S_1}{\partial \xi^A}\mu^A_1]
-\frac{\hbar}{\tau}(\G^{(1)-1},S_1)
+\hbar O(\tau^0)+O(\hbar^2).\label{do}
\ee
Finally, we consider
$\xi^1_{\rho^*}=\exp -\frac{\hbar}{\tau}(\cdot ,
\xi^*\sigma_1\rho^*)\xi=\xi-\frac{\hbar}{\tau}\sigma_1\rho^*$ and
substitute $\xi$ by $\xi^1_{\rho^*}$: 
\be
S_{{R_1}}^{\rho^*}(z,\xi,\rho^*)
\equiv S_{{R_1}}(z,\xi^1_{\rho^*}(\xi,\rho^*))\nonumber\\
=S_{{R_1}}(z,\xi)
-\frac{\hbar}{\tau}\frac{\partial^R S_{{R_1}}}{\partial\xi^A}
\sigma^A_1\rho^*\label{de}\\=S_{{R_1}}(z,\xi)-\frac{\hbar}{\tau}
\frac{\partial^R S_0}{\partial\xi^A}\sigma^A_1\rho^*+\hbar O(\tau^0)
+O(\hbar^2). \label{di}
\ee
We also have that 
\be
\theta^{\rho^*}_{R_1}(z,\xi,\rho^*)\equiv\theta_{R_1}(z,
\xi^1_{\rho^*}(\xi,\rho^*))\nonumber\\
=\theta_{R_1}(z,\xi)
-\frac{\hbar}{\tau}\frac{\partial^R \theta_{{R_1}}}{\partial\xi^A}
\sigma^A_1\rho^*\label{du}\\
=\frac{1}{2\tau}({S^{\rho^*}_{{R_1}}},
{S^{\rho^*}_{{R_1}}})
+\frac{1}{\tau}\Delta_c{S^{\rho^*}_{{R_1}}}\label{da}
.
\ee
Equations (\ref{do}) and (\ref{di}) imply that the action 
\be
{S_{{R_1}}}_{\rho^*}={S^{\rho^*}_{{R_1}}}
+{\theta^{\rho^*}_{{R_1}}}\rho^*,
\ee
with $\tilde\Sigma_1=\Sigma_1-(S_1,\Xi_1)
-\frac{\partial^R S_1}{\partial \xi^A}\mu^A_1$, 
yields a one loop finite effective action 
both in the $\rho^*$ independent and the $\rho^*$ linear part, 
because the terms linear in $\rho^*$ of order $\hbar/\tau$ add up 
precisely  
to $-\theta^{(1)-1}$.
The one loop renormalized and regularized action 
${S_{{R_1}}}_{\rho^*}$ satisfies 
\be
\frac{1}{2}({S_{{R_1}}}_{\rho^*},{S_{{R_1}}}_{\rho^*})+\Delta_c
{S_{{R_1}}}_{\rho^*}=\tau\theta_{{R_1}}^{\rho^*}\nonumber\\=\tau
\frac{\partial^R{S_{{R_1}}}_{\rho^*}}{\partial \rho^*}
+\frac{\partial^R{S_{{R_1}}}_{\rho^*}}
{\partial\xi^B}\hbar\sigma^B_1-\frac{1}{\tau}
\frac{\partial^R{S_{{R_1}}}_{\rho^*}}
{\partial\xi^B}\frac{\partial^R\hbar\sigma^B_1}{\partial\xi^A}\hbar
\sigma^A_1
\rho^*,
\ee
the first equality following from (\ref{da}), and the 
last equality from the expansions (\ref{de}), 
(\ref{du}), together with the identity 
\be
(-)^{B(A+1)}\frac{\partial^R }{\partial \xi^B}
(\frac{\partial^R S_{R_1}}{\partial \xi^A})
\sigma^A_1\sigma^B_1=0. 
\ee 
According to the regularized quantum action principle,
\be
\frac{1}{2}({\G_{{R_1}}}_{\rho^*},{\G_{{R_1}}}_{\rho^*})
+\Delta_c
{\G_{{R_1}}}_{\rho^*}
=\tau
\frac{\partial^R{\G_{{R_1}}}_{\rho^*}}{\partial\rho^*}
+\frac{\partial^R{\G_{{R_1}}}_{\rho^*}}
{\partial\xi^B}\hbar\sigma^B_1\nonumber\\-\frac{1}{\tau}
\frac{\partial^R{\G_{{R_1}}}_{\rho^*}}
{\partial\xi^B}\frac{\partial^R\hbar\sigma^B_1}{\partial\xi^A}
\hbar\sigma^A_1
\rho^*.
\label{hh}
\ee
The $\rho^*$ independent part at one loop and 
lowest order, $\tau^0$, in
$\tau$ gives
\be
\bar s{\G_{{R_1}}}^{(1)0}=\frac{\partial^R
S_0}{\partial\xi^B}\sigma^B_1,
\label{90}
\ee
and shows that only the non trivial part of the anomaly remains.

\subsection{Two loops}

\subsubsection{Equations for the two loop poles}

The one loop renormalized action admits the expansion 
\be
{\G_{R_1}}_{\rho^*}=S_{\rho^*}
+\hbar \Sigma_{n=0}\tau^n{\G_{R_1}}_{\rho^*}^{(1)n}
+\hbar^2\Sigma_{n=-2}\tau^n{\G_{R_1}}_{\rho^*}^{(2)n}
+O(\hbar^3).
\ee
At order $\hbar^2$ (\ref{hh}) gives,
\be
(S_{\rho^*},{\G_{R_1}}^{(2)}_{\rho^*})+\frac{1}{2}
({\G_{R_1}}^{(1)}_{\rho^*},{\G_{R_1}}^{(1)}_{\rho^*})+\Delta_c
{\G_{R_1}}^{(2)}_{\rho^*}\nonumber\\
=\tau\frac{\partial^R{\G_{R_1}}_{\rho^*}^{(2)}}{\partial\rho^*}+
\frac{\partial^R{\G_{R_1}}^{(1)}_{\rho^*}}{\partial\xi^B}\sigma^B_1
-\frac{1}{\tau}\frac{\partial^R S_0}{\partial \xi^B}
\frac{\partial^R\sigma^B_1}{\partial\xi^A}\sigma^A_1
\rho^*.
\label{ht}
\ee
Let ${\G_{R_1}}_{\rho^*}=\G_{R_1}
+\frac{\partial^R{\G_{R_1}}_{\rho^*}}{\partial\rho^*}\rho^*$.
At order $1/\tau^2$, we get, according to the $\rho^*$ independent and
linear parts,
\be
\bar s {\G_{R_1}}^{(2)-2}=0,\label{con1} \\
\bar s(\frac{\partial^R{\G_{R_1}}_{\rho^*}^{(2)-2}}{\partial\rho^*}
-(S_1,{\G_{R_1}}^{(2)-2}))
=0.\label{con2}
\ee
The first of these equations implies:
\begin{lemma}
The second order pole of the 
two loop divergences is a $\bar s$ cocycle.
\end{lemma}
At order $1/\tau$, we get 
\be
\bar s {\G_{R_1}}^{(2)-1}=
\frac{\partial^R{\G_{R_1}}_{\rho^*}^{(2)-2}}{\partial\rho^*}
-(S_1,{\G_{R_1}}^{(2)-2}),\label{ha}\\
\bar s(\frac{\partial^R{\G_{R_1}}_{\rho^*}^{(2)-1}}{\partial\rho^*}
-(S_1,{\G_{R_1}}^{(2)-1})-(S_2,{\G_{R_1}}^{(2)-2}))
=-\frac{\partial^R S_0}{\partial \xi^B}
\frac{\partial^R\sigma^B_1}{\partial\xi^A}\sigma^A_1.\label{gsp}
\ee
Finally, the $\rho^*$ independent part of (\ref{ht}), gives at order
$\tau^0$
\be
\bar s{\G_{R_1}}^{(2)0}+\frac{1}{2}
({\G_{R_1}}^{(1)0},{\G_{R_1}}^{(1)0})=
\frac{\partial^R {\G_{R_1}}^{(1)0}}{\partial \xi^B}\sigma_1^B
\nonumber\\+
\frac{\partial^R{\G_{R_1}}_{\rho^*}^{(2)-1}}{\partial\rho^*}
-(S_1,{\G_{R_1}}^{(2)-1})-(S_2,{\G_{R_1}}^{(2)-2}),\label{gsp1}
\ee 
which allows to identify the combination 
\be
A_2=\frac{\partial^R{\G_{R_1}}_{\rho^*}^{(2)-1}}{\partial\rho^*}
-(S_1,{\G_{R_1}}^{(2)-1})-(S_2,{\G_{R_1}}^{(2)-2})
\ee
as the local
contribution to the two loop anomaly, whereas $\frac{\partial^R
{\G_{R_1}}^{(1)0}}{\partial \xi^B}\sigma_1^B$ is the one loop
renormalized dressing of the non trivial one loop anomaly.

\subsubsection{Two loop anomaly consistency condition}
Before absorbing the divergences, let us consider (\ref{gsp}), which
can be written as 
\be
\bar s A_2=-\frac{1}{2}\frac{\partial^R S_0}{\partial
\xi^B}[\sigma^1,\sigma^1]^B, \label{cons2}
\ee 
where $\xi^*_B[\sigma^1,\sigma^1]^B
\equiv(\xi^*\sigma_1,\xi^*\sigma_1)_\xi$
is an $s_Q$ cocyle because 
of the graded Jacobi identity for the antibracket in $\xi,\xi^*$
space. 
According to equations (\ref{tr1}) and (\ref{tr2}),
this implies that 
\be
\frac{1}{2}(\xi^*\sigma_1,\xi^*\sigma_1)_\xi
= s_Q \xi^*_A\sigma^A_2,\label{999}
\ee
and 
\be
A_2=\bar s  \Sigma_2
+\frac{\partial^R S_0}{\partial \xi^B}\sigma^B_2.\label{100}
\ee

{\bf Discussion:}
We thus see that the consistency condition (\ref{cons2}) on the 
local contribution of the two loop anomaly does not require it to be
just a
cocycle of the extended BRST differential $\bar s$, because of the
non vanishing right hand side. This is in agreement with
the analysis of \cite{PaTr}. Nevertheless, in the extended antifield
formalism, the general solution of (\ref{100}) can be characterized:
it is given by a $\bar s$ boundary up to the term 
$\frac{\partial^R S_0}{\partial \xi^B}\sigma^B_2$
with the following interpretation. From the point of view of
cohomology, equation (\ref{cons2}) should
be understood as a restriction on the non trivial one loop anomalies
$\xi^*_A\sigma^A_1$ that can arise. Indeed, its consequence is
(\ref{999}), which states that the non trivial one loop anomalies
should have a trivial antibracket map\footnote{The antibracket map
here is the antibracket induced in the $s_Q$ cohomology
classes from the antibracket in $\xi$ space. Its construction can be
obtained by just following the construction of the usual antibracket
map used in \cite{BaHe,Bar}.} among themselves. This is a
cohomological statement independent of the choice of representatives.
The term
$\frac{\partial^R S_0}{\partial \xi^B}\sigma^B_2$ of the general
solution for the local part of the two loop anomaly is determined 
by an arbitrary $s_Q$ cocycle up to a
particular solution depending on the choice of representatives for
the non trivial one loop anomalies and needed to make the bracket 
$(\xi^*\sigma_1,\xi^*\sigma_1)_\xi$ $s_Q$ exact. 
This answers, at least in the present context of the extended
antifield formalism and dimensional 
regularization\footnote{The analysis
of \cite{Bar2} confirms these results in the context of algebraic
renormalization.}, 
the question raised in \cite{PaTr} on the
cohomological interpretation of the two loop anomaly consistency
condition. One also sees on this example how the discussion of
the quantum Batalin-Vilkovisky formalism is shifted to $\xi,\xi^*$
space in the extended formalism. 

Note that, as in \cite{Whi}, this result has been achieved by
adding a BRST breaking counterterm, not only for the one loop 
divergences produced
by the standard action itself, but also for the one loop 
divergences produced
by the insertion of the non trivial one loop anomaly. This is
because this anomaly has been coupled to the action itself from the
start, and the
BRST breaking counterterm $\Sigma_1$ also depends on the corresponding
coupling constants.

\subsubsection{Two loop renormalization}

The general solution to (\ref{con1}) is 
${\G_{R_1}}^{(2)-2}=\bar s\Xi_{2,-2}+\frac{\partial^R
S_0}{\partial \xi^A}\mu^A_{2,-2}$.  
We consider the change of variables
\be
z^{2,-2}=\exp -\frac{\hbar^2}{\tau^2}[(\cdot,\Xi^{\rho^*}_{2,-2}) 
+(\cdot,{\xi^1_{\rho^*}}^*
\mu_{2,-2}^{\rho^*})_{\xi^1_{\rho^*}}]
z,\\
{\xi^{2,-2}}
=\exp-\frac{\hbar^2}{\tau^2}(\cdot,{\xi^1_{\rho^*}}^*
\mu_{2,-2}^{\rho^*})_{\xi^1_{\rho^*}}\xi^1_{\rho^*}, 
\ee
where $\Xi^{\rho^*}_{2,-2}(z,\xi,\rho^*)=\Xi_{2,-2}(z,
\xi^1_{\rho^*}(\xi,\rho^*))$ and 
$\mu_{2,-2}^{\rho^*}(\xi,\rho^*)
= \mu_{2,-2}(\xi^1_{\rho^*}(\xi,\rho^*))$.
The fact that we consider this change of variables in terms of
$\xi^1_{\rho^*}$ instead of $\xi$ will not change the
absorption of the $\rho^*$ independent divergences, but it will be
important in order to control the dependence on $\rho^*$ below.
Equation (\ref{ha}) 
means that there is no non trivial part $\frac{\partial^R
S_0}{\partial \xi^A} \sigma^A_{2,-2}$ in the general solution to 
(\ref{con2}) and
hence no need for a renormalization of the coupling constants of order
$\hbar^2/\tau^2$ proportional to $\rho^*$. The general solution to
(\ref{con2}) is 
$\frac{\partial^R{\G_{R_1}}_{\rho^*}^{(2)-2}}{\partial\rho^*}
-(S_1,{\G_{R_1}}^{(2)-2})=\bar
s \Sigma_{2,-2}$, where $\Sigma_{2,-2}$ can be identified with a
particular solution ${\G_{R_1}}_P^{(2)-1}$ of (\ref{ha}).
We take 
\be
{S^{\rho^*}_{R_{2,-2}}}(z,\xi,\rho^*)=
S_{R_1}(z^{2,-2}(z,\xi^1_{\rho^*}),\xi^{2,-2}(\xi^1_{\rho^*}))
+{g_{2,-2}}_A(z,\xi^1_{\rho^*})f^A(\xi^1_{\rho^*})
\nonumber\\-\frac{\hbar^2}{\tau}\tilde
\Sigma_{2,-2}(z^{2,-2}(z,\xi^1_{\rho^*}),
\xi^{2,-2}(\xi^1_{\rho^*})),\nonumber\\
={S_{{R_1}}}-\frac{\hbar^2}{\tau^2}{\G_{R_1}}^{(2)-2}
+O(\hbar^2\tau^{-1})
+O(\hbar^3),
\ee
where $\tilde\Sigma_{2,-2}(z,\xi)$ remains to be determined.
The remark after theorem \ref{t4} again implies
\be
\theta^{\rho^*}_{{R_{2,-2}}}\equiv
\frac{1}{2\tau}({S^{\rho^*}_{R_{2,-2}}},
{S^{\rho^*}_{{R_{2,-2}}}})
+\frac{1}{\tau}\Delta_c{S^{\rho^*}_{{R_{2,-2}}}}
\nonumber\\
={\theta^{\rho^*}_{{R_1}}}-\frac{\hbar^2}{\tau^2}\bar s
[\tilde\Sigma_{2,-2}+\frac{\partial^R S_1}{\partial
\xi^A}\mu^A_{2,-2}+(S_1,\Xi_{2,-2})]\nonumber\\
-\frac{\hbar^2}{\tau^2}(S_1,{\G_{R_1}}^{(2)-2})+\hbar^2 O(\tau^{-1})
+O(\hbar^3).
\ee
The action 
\be
{S_{{R_{2,-2}}}}_{\rho^*}={S^{\rho^*}_{{R_{2,-2}}}}
+{\theta^{\rho^*}_{{R_{2,-2}}}}\rho^*,
\ee
with $\tilde\Sigma_{2,-2}=\Sigma_{2,-2}-\frac{\partial^R S_1}{\partial
\xi^A}\mu^A_{2,-2}+(S_1,\Xi_{2,-2})$, yields an effective action 
${\G_{{R_{2,-2}}}}_{\rho^*}$
without $\hbar^2/\tau^2$ divergences and only simple poles at order
$\hbar^2$, because the terms linear in $\rho^*$ of order 
$\hbar^2/\tau^2$ add up precisely to
$-\frac{\partial^R{\G_{R_1}}_{\rho^*}^{(2)-2}}{\partial\rho^*}$.
We have again that 
\be
\frac{1}{2}({S_{{R_{2,-2}}}}_{\rho^*},{S_{{R_{2,-2}}}}_{\rho^*})
+\Delta_c{S_{{R_{2,-2}}}}_{\rho^*}
=\tau{\theta^{\rho^*}_{{R_{2,-2}}}}\nonumber\\
=\tau\frac{\partial^R{S_{{R_{2,-2}}}}_{\rho^*}}{\partial\rho^*}
+\frac{\partial^R{S_{{R_{2,-2}}}}_{\rho^*}}{\partial\xi^B}
\hbar\sigma^B_1
-\frac{1}{\tau}
\frac{\partial^R{S_{{R_{2,-2}}}}_{\rho^*}}
{\partial\xi^B}\frac{1}{2}[\hbar\sigma_1,
\hbar\sigma_1]^B
\rho^*.
\ee
The last equation follows from the fact that the dependence of
$S^{\rho^*}_{{R_{2,-2}}}$ and ${\theta^{\rho^*}_{{R_{2,-2}}}}$ 
on $\rho^*$ is, as before, 
through the combination $\xi^1_{\rho^*}$.
The same equation holds again for the effective action:
\be
\frac{1}{2}({\G_{{R_{2,-2}}}}_{\rho^*},{\G_{{R_{2,-2}}}}_{\rho^*})
+\Delta_c{\G_{{R_{2,-2}}}}_{\rho^*}
=\tau\frac{\partial^R{\G_{{R_{2,-2}}}}_{\rho^*}}{\partial\rho^*}
+\frac{\partial^R{\G_{{R_{2,-2}}}}_{\rho^*}}{\partial\xi^B}
\hbar\sigma^B_1
\nonumber\\-\frac{1}{\tau}
\frac{\partial^R{\G_{{R_{2,-2}}}}_{\rho^*}}
{\partial\xi^B}\frac{1}{2}[\hbar\sigma_1,
\hbar\sigma_1]^B
\rho^*.
\ee
The expansion of this effective action is 
\be
{\G_{R_{2,-2}}}_{\rho^*}=S_{\rho^*}
+\hbar \Sigma_{n=0}\tau^n{\G_{R_{2,-2}}}_{\rho^*}^{(1)n}
+\hbar^2\Sigma_{n=-1}\tau^n{\G_{R_{2,-2}}}_{\rho^*}^{(2)n}
+O(\hbar^3).
\ee

The divergences $\G_{{R_{2,-2}}}^{(2)-1}$ and
$\frac{\partial^R{\G_{R_{2,-2}}}_{\rho^*}^{(2)-1}}{\partial\rho^*}$, 
now satisfy $\bar s
\G_{{R_{2,-2}}}^{(2)-1}=0$ and $\bar s A^\prime_2
=
\frac{\partial^R S_0}{\partial
\xi^A}\frac{1}{2}[\sigma^1,\sigma^1]^A$, with $A^\prime_2=
\frac{\partial^R{\G_{R_{2,-2}}}_{\rho^*}^{(2)-1}}{\partial\rho^*}-
(\G_{{R_{2,-2}}}^{(2)-1},S_1)$. The general solutions are 
$\G_{{R_{2,-2}}}^{(2)-1}=\bar s \Xi_{2,-1}
+\frac{\partial^R S_0}{\partial
\xi^A}\mu_{2,-1}$ and $A^\prime_2=\bar s\Sigma_{2,-1}
+\frac{\partial^R S_0}{\partial
\xi^A}\sigma_2^A$. As in the one loop case, one first subtracts a
suitably defined BRST breaking counterterm, then one makes the
field-antifield 
and coupling constant redefinition determined by 
$\Xi^{\rho^*}_{2,-1}$ and
$\mu^{\rho^*}_{2,-1}$, and finally, one substitues 
$\xi^1_{\rho^*}$ everywhere by
$\xi^2_{\rho^*}=\xi^1_{\rho^*}-\frac{\hbar^2}{\tau}\sigma_2\rho^*$, 
giving a total $\rho^*$ dependence through the combination 
$\xi^2_{\rho^*}=\xi
-\frac{\hbar}{\tau}\sigma_1\rho^*
-\frac{\hbar^2}{\tau}\sigma_2\rho^*$. 

Using the same arguments as in the one loop case, one finally finds
that the two loop renormalized and regularized action
${S_{R_2}}_{\rho^*}$ satisfies 
\be
\frac{1}{2}({S_{{R_{2}}}}_{\rho^*},{S_{{R_{2}}}}_{\rho^*})
+\Delta_c{S_{{R_{2}}}}_{\rho^*}
=\tau
\frac{\partial^R{S_{{R_{2}}}}_{\rho^*}}{\partial\rho^*}
+\frac{\partial^R{S_{{R_{2}}}}_{\rho^*}}{\partial\xi^B}
(\hbar\sigma^B_1+\hbar^2\sigma^B_{2})\nonumber\\
-\frac{1}{\tau}\frac{\partial^R{S_{{R_{2}}}}_{\rho^*}}{\partial\xi^B}
\frac{1}{2}[\hbar\sigma_1+\hbar^2\sigma_{2},
\hbar\sigma_1+\hbar^2\sigma_{2}]^B)\rho^*,
\ee
the same equation holding for the 
two loop renormalized effective action 
${\G_{{R_{2}}}}_{\rho^*}$.

\subsection{Higher orders}

It is then possible to continue recursively to higher loops 
to get a completely subtracted and regularized action 
${S_{{R_{\infty}}}}$. It is obtained from 
\be
S_\tau-\sum_{n=1}\frac{\hbar^n}{\tau^{n-1}}\sum_{k=0}^{n-1}\tau^k
\tilde\Sigma_{n,k-n},
\ee
with suitably choosen BRST breaking counterterms 
$\tilde\Sigma_{n,k-n}$,
by successive canonical field-antifield and 
coupling constants redefinitions.  
It satisfies
\be
\frac{1}{2}({S_{{R_\infty}}}_{\rho^*},{S_{{R_\infty}}}_{\rho^*})
+\Delta_c{S_{{R_\infty}}}_{\rho^*}
=\tau\frac{\partial^R{S_{{R_\infty}}}_{\rho^*}}{\partial\rho^*}+
\frac{\partial^R{S_{{R_\infty}}}_{\rho^*}}{\partial\xi^B}
\sigma^B\nonumber\\
-\frac{1}{\tau} 
\frac{\partial^R{S_{{R_\infty}}}_{\rho^*}}{\partial\xi^B}
\frac{1}{2}[\sigma,\sigma]^B\rho^*.
\label{end1}
\ee
The corresponding 
completely renormalized and regularized effective action 
${\G_{{R_\infty}}}_{\rho^*}$ satisfies the same equation. 
\be
\frac{1}{2}({\G_{{R_\infty}}}_{\rho^*},{\G_{{R_\infty}}}_{\rho^*})
+\Delta_c{\G_{{R_\infty}}}_{\rho^*}
=\tau\frac{\partial^R{\G_{{R_\infty}}}_{\rho^*}}{\partial\rho^*}+
\frac{\partial^R{\G_{{R_\infty}}}_{\rho^*}}{\partial\xi^B}
\sigma^B\nonumber\\
-\frac{1}{\tau} 
\frac{\partial^R{\G_{{R_\infty}}}_{\rho^*}}{\partial\xi^B}
\frac{1}{2}[\sigma,\sigma]^B\rho^*.\label{end}
\ee
One can then
put $\rho^*=0$ and take safely the limit $\tau\longrightarrow 0$,
because there are no more divergences left. The renormalized 
effective action 
$\Gamma^R=\lim_{\tau\rightarrow
0}{\G_{{R_\infty}}}_{\rho^*}|_{\rho^*=0}$ satisfies 
\be
\frac{1}{2}(\G^R,\G^R)
+\Delta_c\G^R
=\frac{\partial^R\G^R}{\partial\xi^B}
\sigma^B.
\ee
\begin{theorem}
The absorption of the divergences of a theory 
in dimensional renormalization involves, besides
redefinitions of the solution of the extended master equation,
determined by anticanonical field-antifield and coupling constant 
renormalizations, only the subtraction of suitably chosen BRST breaking
counterterms. The renormalization can be done in 
such a way that the anomalous breaking of the extended Zinn-Justin
equation is determined to all orders by the cohomology $H^1(\bar s,F)$.
\end{theorem}

The corresponding result for the anomalies 
in the standard Batalin-Vilkovisky formalism 
has been obtained in the framework of algebraic renormalization 
in \cite{Whi}.

\subsection{The quantum Batalin-Vilkovisky $\Delta$ operator}

In \cite{TNP,PaTr}, explicit expression for the $\Delta$
operator have been obtained in the context of Pauli-Villars and non
local regularization respectively. The aim of this section is to get
such an expression in the context of dimensional renormalization. 
The expression we will get here will be defined on all the
generalized observables of the theory, and not only on $S$ alone,
since they are contained in the solution $S(\xi)$ of the extended
master equation. 

As discussed for instance in section 4 of \cite{DPT} in the context of
the BPHZ renormalized antifield formalism, even though there is a well
defined expression for the anomaly, there is no room for the
formal Batalin-Vilkovisky $\Delta$ operator in the final renormalized
theory. 
Contact with the quantum Batalin-Vilkovisky formalism in the present 
set-up has thus to be done on the renormalized theory before the
regulator $\tau$ is removed. Moreover, as in the previous 
discussion of the renormalization, it turns out to 
be important not to put to zero the
fermionic variable $\rho^*$, which couples the breaking of the extended
master equation due to the regularization.
Let us introduce the notation 
$W={S_{R_\infty}}_{\rho^*}$ for
the completely renormalized and regularized action
and ${\cal A}=\frac{\partial^R W}{\partial
\xi^B}(\sigma^B-\frac{1}{2\tau}[\sigma,\sigma]^B\rho^*)$
so that (\ref{end}) can be
written,
\be
\frac{1}{2}({\G_{{R_\infty}}}_{\rho^*},{\G_{{R_\infty}}}_{\rho^*})
+\Delta_c{\G_{{R_\infty}}}_{\rho^*}+
\tau\frac{\partial^L{\G_{{R_\infty}}}_{\rho^*}}{\partial\rho^*}
={\cal A}\circ{\G_{{R_\infty}}}_{\rho^*},\label{blo}
\ee
while (\ref{end1}) becomes,
\be
\frac{1}{2}(W,W)
+\Delta_c W
+\tau\frac{\partial^L W}{\partial\rho^*}
={\cal A}.\label{bla}
\ee
Let us define $\Delta_d=\frac{\tau}{-i\hbar}\frac{\partial^L
}{\partial\rho^*}$, so that we can write the above equation as 
\be
\frac{1}{2}(W,W)
+\Delta_c W-i\hbar\Delta_d W={\cal A}.\label{blu}
\ee
The operator $\Delta_d$ is of ghost number $1$, it is nilpotent,
$\Delta_d^2=0$, it anticommutes with $\Delta_c$, 
$\{\Delta_c,\Delta_d\}=0$, and it is a graded
derivation of the antibracket, i.e., it satisfies equation (\ref{der})
(with $\Delta_c$ replaced by $\Delta_d$). 
Using the properties of the antibracket, $\Delta_c$ and $\Delta_d$, it
follows that, by applying $(W,\cdot)+\Delta_c-i\hbar\Delta_d$ to
(\ref{blu}), the left hand side vanshes identically. This gives the
consistency condition
\be
(W,{\cal A})+\Delta_c{\cal A}-i\hbar\Delta_d{\cal A}=0.
\ee

{\bf Discussion:}
Starting from the path integral expression
\be
Z(J,\phi^*,\xi,\rho^*)=\int {\cal D}\phi
\exp\left(\frac{i}{\hbar}[W+\int d^nx\ J_A\phi^A]\right),
\ee
with associated effective action ${\G_{{R_\infty}}}_{\rho^*}$,
standard formal path integral manipulations using integrations 
by parts 
give
\be
\frac{1}{2}({\G_{{R_\infty}}}_{\rho^*},{\G_{{R_\infty}}}_{\rho^*})
+\Delta_c{\G_{{R_\infty}}}_{\rho^*}={\cal A}^\prime\circ
{\G_{{R_\infty}}}_{\rho^*},
\ee
where 
\be
{\cal A}^\prime=\frac{1}{2}(W,W)+\Delta_c
W-i\hbar''\Delta W''.
\ee
This expression involves the second order functional 
derivative operator 
$\Delta=(-)^{A+1}\frac{\delta^R}{\delta\phi^A(x)}
\frac{\delta^R}{\delta\phi^*_A(x)}$.
The quotation marks mean that the above definition of $\Delta$ cannot
be used since $\Delta$ is ill defined when acting on local functionals
and thus on $W$. Using (\ref{blo}) for the left hand side, we get 
${\cal A}^\prime=-\tau\frac{\partial^L W}{\partial\rho^*}+{\cal
A}$. Using furthermore (\ref{bla}), it follows that $-i\hbar''\Delta
W''=0$, as was to be expected in dimensional regularization, where
$''\delta(0)''=0$. 

In equation (\ref{bla}), obtained by an analysis of the
renormalization procedure, there appears the operator 
$\Delta_d$, which is unexpected from the
point of view of formal path integral manipulations, not taking the
regularization and renormalization into account. Furthermore, the 
operator $\Delta_d$ has the same algebraic properties as the formal
operator  $\Delta$, when acting on local functionals.
In dimensional regularization, one has traded the operator
$\Delta$, vanishing on local functionals, for the operator
$\Delta_d$. We thus find, in the context of dimensional
regularization, that the role of the Batalin-Vilkovisky $\Delta$
operator is played by the operator $\Delta_d$, 
introduced originally in the last reference of \cite{Ton}. 

Furthermore, (\ref{bla}) suggests that the operator $\Delta_c$, 
can be understood as a classical part of the 
Batalin-Vilkovisky $\Delta$
operator in the extended antifield formalism. This interpretation is
supported by the fact that both
$\Delta_c$ and $\Delta_d$ arise in a similar way from an extended 
action satisfying a standard master equation in an extended space with
an enlarged bracket: this was shown for $\Delta_c$ in section 1.4.  
In \cite{Ton1} in the context of the standard Batalin-Vilkovisky 
formalism, it was shown that $\Delta_d$ also arises
from an ``improved'' classical
master equation, if 
the space of fields and
antifields is enlarged to enclude the global pair of 
variables $\rho,\rho^*$, the antibracket is 
extended to this pair and the
regularized action is extended to $S_\tau+\theta_\tau\rho^*+\tau\rho$. 

It might be worthwhile to point out that the there are two other
``extended master equations'' which can be understood in this way. 
(i) The
famous tree level Slavnov-Taylor identity in Yang-Mills theory,
gauge fixed with the help of the auxiliary $B$ field
\cite{BRS,Zin}, can be obtained from the antifield formalism 
(before gauge fixation) by adding to
the minimal solution of the master equation the term $\int d^nx \bar
C^*_a B^a$ and extending the antibracket to the pair $\bar C^a,B^a$ and
their antifields.
(ii) The extended Slavnonv-Taylor identity \cite{PiSi} including the
BRS doublet $\alpha$ and $\chi$ introduced in order to control the
dependence of the theory on the parameter $\alpha$ of covariant linear
gauges can be obtained in the antifield formalism by introducing the
global pair $\alpha,\chi$ and their antifields and adding the term
$\chi^*\alpha$ to the solution of the master equation \cite{Fis}.

\section{Conclusion}

In order to study the renormalization of a theory together with
all its non trivial observables (described in the
Batalin-Vilkovisky formalism by the BRST cohomology classes), it is 
natural \cite{KlZu} to couple these observables (more precisely, those 
that are not already present in the solution to the master equation
coupled through essential coupling constants) with the help of 
new coupling
constants. Considering such an action as a starting point for
renormalization theory allows on the one hand to get information on
the renormalization of the operators, but also \cite{Bar} constraints
on the divergences of the starting point theory without the additional
operators. The action plus the additional observables satisfies the
master equation in general only up to second order in the new
coupling constants since the cohomology classes may depend on the
antifields. 

In this paper, we have constructed an extension of the
standard solution satisfying an extended master
equation. This extended master equation 
governs not only the symmetries of the starting point action, but also 
the antibracket algebra of the observables. It is the right tool to study
renormalizability of a theory compatible with symmetries for the
following reasons:

(i) the divergences of the
theory are constrained by the differential associated to this
extended master equation, 

(ii) the absorption of these divergences
can be performed order by order through  
redefinitions of the extended action determined by 
field-antifield and coupling constant redefinitions in such a way that the
redefined action still satisfies the extended master equation, up to
breakings due to the non invariant terms $\sum_{n=1}\tau^n S_n$ 
determined by the regularization and associated symmetry breaking
counterterms $\Sigma_{n,k-n}$, 

(iii) the associated renormalized 
effective action satisfies the extended master
equation up to non trivial 
anomalous breakings determined by the cohomology of the
extended BRST differential in ghost number $1$.

If one adapts the terminology of \cite{GoWe} 
and defines a theory to be renormalizable in the 
modern sense if properties (ii) and (iii) hold, our
results mean that all theories are renormalizable in this
sense. 

The approach proposed in this paper is not
completely formal, since on the one hand, 
the higher order maps can be
computed in principle in a straightforward way 
once the local BRST cohomology
groups of the theory are known, and on the other hand, there has been
considerable progress in computing these groups for various
models such as Yang-Mills and Chern-Simons theory
\cite{BBH}, gravity \cite{BBH2,BTV}, $p$-form gauge theories
\cite{HKS}, $N=1$ supergravity \cite{Bra} or $D$-strings \cite{BGS}.  

For a theory, where the BRST cohomology and the higher order maps are 
completely known, the only remaining problem consists in computing the
$s_Q$ cohomology. 

\section*{Acknowledgments}

The author thanks F.~Brandt, C.~Schomblond, J.~Gomis, P.A.~Grassi, 
M.~Henneaux, T.~Hurth, J.~Stasheff, M.~Tytgat and A.~Wilch for 
useful discussions.  

\section*{Appendix: Elimination of antifield dependent counterterms 
in Yang-Mills theories with $U(1)$ factors}

In this appendix, we will discuss the elimination by higher order
cohomological restrictions of a type of 
antifield dependent counterterms arising in non
semi-simple Yang-Mills theories. They have been discussed for the
first time in detail in 
\cite{BBBC}, were analyzed from a cohomological point of view 
in \cite{BBH}
and reconsidered in the concrete context of the standard
model in \cite{Gra}. These counterterms 
(or instabilities in the terminology of \cite{BBBC,Gra}) have the
following general structure \cite{BBH}: 
$$
K^\prime=f^\Delta_\alpha\int d^nx\ j^\mu_\Delta A_\mu^\alpha
+(A^{*\mu}_a
X^a_{\mu\Delta}+y^*_iX^i_\Delta)C^\alpha,
$$ 
where $f^\Delta_\alpha$ are constants, $A_\mu^\alpha$ abelian gauge
fields, $j^\mu_\Delta$ non trivial conserved currents and
$\delta_\Delta A^a_\mu=X^a_{\mu\Delta},\delta_\Delta y^i=X^i_\Delta$ 
the generators of the corresponding
symmetries on all the gauge fields $A_\mu^a$ and the matter fields
$y^i$. 
In order to eliminate these
instabilities by cohomological means, we will show that:

\noindent{\it it is
sufficient that there exists a set
of local, non integrated, off-shell gauge invariant polynomials $O_\Gamma(x)$
constructed out of the $A^a_\mu,y^i$ and their derivatives,
that break the global symmetries $\delta_\Delta$  
in the following sense: the variation of $O_\Gamma(x)$ under the gauged
global symmetries $\delta_\Delta$ with gauge parameter given by
$f^\Delta_\alpha \epsilon^\alpha$ should not be equal on shell 
to an ordinary gauge transformation (involving the abelian gauge
parameters $\epsilon^\alpha$ alone) of some local polynomials 
$P_\Gamma(x)$ constructed 
out of the $A^a_\mu,y^i$ and their derivatives.} 

Indeed,
using the
extended action
\footnote{The author thanks P.A. Grassi for suggesting the use of
external sources in this example.} 
$S_{k(x)}=S+\int d^nx\ k^\Gamma(x) O_\Gamma(x)$, which
satisfies $1/2(S_{k(x)},S_{k(x)})=0$ and the corresponding regularized
action principle, it follows from the equation independent of the sources
$k(x)$ that the divergences ${\G^{(1)}}_{div}$ of the theory without
$k(x)$ are, as usual, required to be BRST invariant. 
The terms linear in $k(x)$ then imply
\be
(\left.\frac{\delta {\G^{(1)}_{k(x)}}_{div}}{\delta
k^\Gamma(x)}\right|_{k(x)=0},S)+(O_\Gamma(x),{\G^{(1)}}_{div})=0.
\label{sudu}
\ee
The second term of this equation gives
for the antifield dependent counterterms 
${\G^{(1)}}_{div}=K^\prime$
above $(O_\Gamma(x),K^\prime)=(C^\alpha f^\Delta_\alpha
\delta_{\Delta}) O_\Gamma(x)$, 
because we have choosen (for simplicity) $O_\Gamma(x)$ to be
independent of the antifields. From (\ref{sudu}), it then follows that 
$(C^\alpha f^\Delta_\alpha \delta_{\Delta}) O_\Gamma(x)$ 
must be given on-shell by a 
gauge transformation,
involving the abelian ghosts alone, of polynomials $P_\Gamma(x)$. 
This follows
by using the explicit form of the BRST differential, and after
evaluation, putting to zero the antifields, and the non abelian
ghosts. Hence, the counterterms $K^\prime$ are excluded a priori 
whenever it is possibe
to construct $O_\Gamma(x)$'s for which the corresponding $P_\Gamma(x)$
do not exist so that (\ref{sudu}) cannot be satisfied.

{\bf Remark}: Because we use external sources instead
of coupling constant, the example given here is not covered by the
general analysis done in this paper. Note however that it is
possible to generalize the antibracket map 
considered in \cite{BaHe,Bar} to a mixed antibracket map from 
the tensor product of 
integrated times non integrated cohomology classes to non 
integrated cohomology classes as needed in this example. The invariant
cohomological statement of which we have discussed a particular case 
above is: {\it
the counterterms are restricted to belong to the kernel of the mixed
antibracket map}. We use
external sources in this example because the restrictions we get are
stronger and the discussion is simplified: we need not worry 
about possible integrations by parts (in
momentum space, this means that the restrictions we get are valid for
all values of the external momentum and not only for zero external
momentum). 

This means that besides the arguments of \cite{BBBC,GoWe,Gra}, there
exists an elegant cohomological mechanism to eliminate this
type of antifield dependent counterterms.
 
In the concrete case of the standard model, the global symmetries
$\delta_\Delta$ correspond to lepton and baryon number
conservation. There is only one abelian ghost $C^\alpha$, the abelian gauge
transformation of the matter fields being $\delta_{abelian} y=i{\cal
Y}y$, where ${\cal Y}={\cal Y}^i_jy^j{\partial\over\partial y^i}$ 
is the hypercharge. As an example of $O_\Gamma$'s we can take any three
linearily independent operators out of the lepton number non
conserving gauge invariant operators of dimension $5$ in the
matter fields given in eq.(20) of \cite{Wei3}
(they can also be found in eq. (21.3.54) of \cite{Wei4}) and one
baryon number non conserving operator out of the six dimension $6$
gauge invariant operators given in eqs. (1)-(6) in \cite{Wei3,WiZe}.
Because these operators are build out of the undifferentiated matter
fields alone, a sufficient
condition for (\ref{sudu}) to hold is the existence of
$P^\prime_{\Gamma}(x)$'s build out of the undifferentiated $y^i$ such
that 
\be
f^\Gamma n_\Gamma O_\Gamma={\cal Y}P^\prime_{\Gamma},\label{ult}
\ee
(with no summation over $\Gamma$), 
where $n_\Gamma$ is the lepton number of the $O_\Gamma$'s for 
$\Gamma=1,2,3$ and the
baryon number for $O_4$.
This follows by identifying the term in the abelian ghost and putting, in 
addition to the non abelian ghosts and the antifields, 
the derivatives of the abelian ghost, the derivatives
of the matter fields and all the gauge fields to zero and using the fact
that the equations of motion necessarily involve derivatives. 
Because the $O_\Gamma$'s we have choosen
are all of homogeneity $4$ in the $y^i$ and ${\cal Y}$ is of
homogeneity $0$, we can assume that the homogeneity of the 
$P^\prime_{\Gamma}$'s is also $4$. By decomposing the space $M_4$ of
monomials of homogenity $4$ in the $y^i$ into eigenspaces of the
hermitian operator ${\cal Y}$ with definite eigenvalues
$M_4=M_4^0+\oplus_{n \neq 0} M_4^n$, it follows that (\ref{ult}) has
no non trivial solutions. Indeed, decomposing 
$P^\prime_{\Gamma}=P^{0\prime}_{\Gamma}+
\sum_{n\neq 0} P^{n\prime}_{\Gamma}$, (\ref{ult})
reads $f^\Gamma n_\Gamma O_\Gamma=\sum_{n\neq
0}n P^{n\prime}_{\Gamma}$. Applying ${\cal Y}$ $k$ times and using the
fact that gauge invariance of $O_\Gamma$ implies ${\cal
Y}O_{\Gamma}=0$, we get $\sum_{n\neq
0}n^k P^{n\prime}_{\Gamma}=0$. We then can conclude that
$P^{n\prime}_{\Gamma}=0$ for $n\neq 0$, which implies $f^\Gamma=0$. 

As usual, this one loop reasoning can be extended recursively to
higher orders, or alternatively, it can be 
discussed independently of the assumption that there exists 
an invariant regularization scheme in the context of
algebraic renormalization.

\end{document}